\documentclass[reprint,amsmath,amssymb,aps,prb,longbibliography,showpacks]{revtex4-2}

\usepackage{graphicx}% Include figure files
\usepackage{dcolumn}% Align table columns on decimal point
\usepackage{bm}% bold math
\usepackage{hyperref}% add hypertext capabilities

\usepackage{color}

\usepackage[T2A]{fontenc}
\usepackage[utf8]{inputenc}
\usepackage[english]{babel}
\usepackage[dvipsnames]{xcolor}
\usepackage{svg}
\usepackage{amssymb}
\usepackage{amsmath}
\usepackage{amsfonts}
\usepackage{appendix}
\usepackage{braket}
\usepackage{multirow}
\usepackage{natbib}
\usepackage[normalem]{ulem}
\usepackage{verbatim}
\usepackage{dcolumn}
\usepackage{ulem}
\usepackage{float}
\usepackage{latexsym}
\usepackage{dcolumn}
\usepackage{amsmath}
\usepackage{epsf}

\usepackage{soul}

\usepackage[normalem]{ulem}

\def \be {\begin{equation}}
\def \ee {\end{equation}}
\def \bea {\begin{align}}
\def \eea {\end{align}}

\def \BEA {\begin{eqnarray}}
\def \EEA {\end{eqnarray}}

\def \BC {\begin{cases}}
\def \EC {\end{cases}}

\begin{document}

\preprint{APS/123-QED}

\title
{Tunable massive and acoustic plasmons in two-dimensional plasmonic crystals }

\author{I.\,V.\,Gorbenko$^{a}$, P.\,A.\,Gusikhin$^{b}$, V.\,Yu.\,Kachorovskii$^{a}$, V.\,M.\,Muravev$^{b}$}
\affiliation{$^a$ Centre of Nanoheterostructures, Ioffe Institute, 194021 St. Petersburg, Russia \\
$^b$ Osipyan Institute of Solid State Physics, RAS, Chernogolovka, 142432 Russia
}

\date{\today}

\begin{abstract}
We theoretically investigate dispersion of plasma waves propagating in a lateral plasmonic crystal based on a two-dimensional electron system with grating gates. Two specific configurations are analyzed: a system with single grating gate having ungated gaps and a double-grating-gate system. We calculate the dispersion relations for the fundamental and several higher-order plasma modes, classifying them as either {\it bright} or {\it dark} excitations. At the boundaries of the Brillouin zones, the dispersion of both types of excitations is shown to be quadratic, justifying introduction of effective bright and dark plasmon masses. In the low-frequency limit, the plasmonic crystal spectrum exhibits an acoustic plasma mode characterized by a certain velocity. We demonstrate that the effective plasmon mass and acoustic velocity are highly sensitive to both the crystal geometry (specifically the lattice filling factor) and the gate voltages, enabling wide-range tunability.
\end{abstract}

\maketitle

\section{Introduction}

Plasma oscillations in low-dimensional systems have been intensively investigated in recent decades, offering both fundamental insights and significant potential for applications~\cite{Maier2007}. A particularly promising and rapidly advancing field within plasmonics is the physics of lateral plasmonic crystals (LPC). These systems comprise a two-dimensional electron system (2DES) coupled to a periodic metal grating gate or a double-grating gate structure. A key feature is that the plasma wave velocity in such systems can be spatially modulated and electrically tuned over a wide range by applying the gate voltage or two voltages in the case of double-grating system. For micron-scale gratings, the plasma oscillation frequency lies within the terahertz (THz) range, which is highly relevant for device applications. This prospect for THz electronics has motivated extensive research,
 including the development of tunable THz generators~\cite{Dyakonov1993, Dyakonov1996, Mikhailov1998, Kachorovskii2012, Moiseenko2020, Boubanga2020}, detectors~\cite{Allen2002, Knap2009, Muravev2012, Lusakowski2014, Gayduchenko2021}, phase shifters~\cite{Muravev2022, Muravev2023}, and current rectifiers based on the ``ratchet'' effect~\cite{Ivchenko2011, Popov2011, Olbrich2011, Rozhansky2015, Olbrich2016, Faltermeier2017, Faltermeier2018, Hubmann2020, Sai2021, Monch2022, Monch2023}.

In the pioneering experiments investigating plasma excitations in two-dimensional semiconductor systems, a metal grating with period $L$ served as a coupler to excite plasmons with a wave vector $q=2\pi/L$~\cite{Allen1977, Theis1977, Tsui1978, Theis1978, Theis1980, Tsui1980a, Heitmann1982, Kotthaus1988}. Interest in plasmonic crystals -- characterized by the emergence of allowed and forbidden bands for propagating two-dimensional plasmons -- emerged much later, driven in part by advances in semiconductor nanofabrication. The primary advantage of plasmonic crystals is the ability to engineer and control their band structure. Experimental tunability has already been demonstrated using gate-voltage control~\cite{Muravjov2010, Boubanga2020, Sai2023, Dub2024}, applying an external magnetic field~\cite{Muravev2008}, and by changing the grating-gate geometry~\cite{Khisameeva2025}. 
Lateral plasmonic crystals are currently the subject of intensive theoretical and numerical study~\cite{Mikhailov1998, Fateev2010, Kachorovskii2012, Aizin2012, Popov2015, Petrov2017, Fateev2019a, Moiseenko2020, Aizin2023, Aizin2024, Gorbenko2024, Gorbenko2025}, complemented by experimental investigations~\cite{Mackens1984, Otsuji2008, Muravjov2010, Ju2011, Dyer2012, Otsuji2013EmissionGraphene, Dyer2013, Boubanga2020, Sai2023, Dub2024, Khisameeva2025}.
It is also worth noting that plasmon resonances with relatively high quality factors have been observed at elevated temperatures: up to $170$~K in AlGaN/GaN-based structures~\cite{Muravjov2010} and up to room temperature in graphene-based structures~\cite{Boubanga2020}.

Despite substantial progress in understanding plasmonic-crystal dispersion, several important issues remain unresolved. So far, essentially all experiments have been performed in a geometry where the incident electromagnetic wave is normal to the plane of the crystal’s 2DES. Consequently, the excitation is restricted to discrete wave vectors $q=N \times 2 \pi/L$ (with $N= 1, 2, 3 \ldots$) in the extended Brillouin zone. The dispersion away from these isolated points has received little attention. 

In this paper, we present a detailed analysis of the dispersion of the fundamental and several higher-order two-dimensional plasma waves propagating in a plasmonic crystal. We distinguish the so-called {\it bright} and {\it dark} modes~\cite{Gorbenko2024}. We show that, at the edges of the Brillouin zones, the dispersion of both bright and dark plasma excitations is quadratic. This behavior justifies the introduction of effective bright and bright plasmon masses, denoted as $m_{\rm b}$ and $m_{\rm d}$, respectively. Our analysis further reveals that these effective masses are highly sensitive to the system geometry (specifically the filling factor of the plasmonic crystal lattice) and can be tuned across a broad range via the gate-grid voltage. We also show that, in the low-frequency limit, an acoustic plasma wave with a linear dispersion can propagate in a plasmonic crystal, with an average velocity that is tunable via the gate. Linear-dispersion plasmon modes can also emerge at higher frequencies 
for a specific gate voltage values that satisfy the condition $m_{\rm b} = m_{\rm d} = 0.$
Finally, we analyze the spatial localization of plasmons in the crystal and how it can be controlled by the gate voltage.

\section{Theoretical Model}

\begin{figure}[t!]
\centering
\includegraphics[width=0.9 \linewidth]{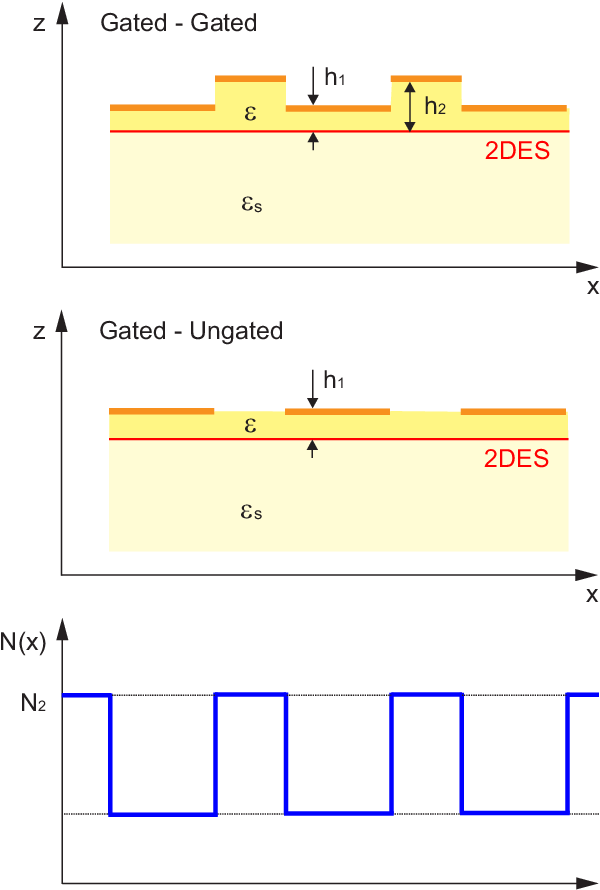}
\caption{Schematic depiction of gated-gated (top) and gated-ungated (middle) structures, and the model dependence of electron concentration $N$ on $x$ (bottom) within an elementary cell of the plasmonic crystal (consisting of regions $L_1$ and $L_2$). The 2D electron gas (2DEG in the figure) occupies the $(x, y)$-plane, is homogeneous in the $y$-direction, and is periodically modulated in the $x$-direction.
}
\label{Fig-grating}
\end{figure}

For the description of two-dimensional plasma waves in a plasmonic crystal, we use the Kronig-Penney model~\cite{Kachorovskii2012, Popov2015, Petrov2017, Boubanga2020, Gorbenko2024, Khisameeva2025}, the simplest model that allows to quantitatively describe plasma waves in periodically modulated systems. In this model, the unit cell of length $L$ is partitioned into two regions with different properties. The first implementation comprises two gated sections with metallic gates of lengths $L_{\rm g1}$ and $L_{\rm g2}$ satisfying
$L=L_{\rm g1}+L_{\rm g2}$ (Fig.~\ref{Fig-grating}(a)). The second implementation consists of gates of length $L_{\rm g}$ separated by ungated gaps of length $L_{\rm u}$, so that $L=L_{\rm g}+L_{\rm u}$ (Fig.~\ref{Fig-grating}(b)). In what follows, we analyze both plasmonic crystal geometries.

To parameterize the plasmonic crystal unit-cell configuration, we introduce the parameter:
\be
f=\left\{
\begin{array}{ll}
    &  \displaystyle \frac{L_{\rm g1}}{L_{\rm g1}+L_{\rm g2}},\quad{\rm for ~gated-gated ~case,} 
    \\
          & \displaystyle \frac{L_{\rm g}}{L_{\rm g}+L_{\rm u}},\quad{\rm for ~gated-ungated ~case,}
    \end{array}
\right.
\ee
% \be
% f= \frac{L_{\rm g1}}{L_{\rm g1}+L_{\rm g2}}, \qquad f= \frac{L_{\rm g}}{L_{\rm g}+L_{\rm u}}
% \ee
which, in the gated–ungated geometry, has the natural interpretation of the filling fraction of the lattice occupied by gated regions.

We consider two classes of plasma waves in a two-dimensional electron system (2DES): screened and unscreened. In both cases, the long-wavelength plasmon dispersion $\omega_p(q)$ is given by~\cite{Stern1967, Chaplik1972}
\begin{equation}
\omega_p = \sqrt{\frac{ 2 \pi N e^2 }{ m^{\ast} \, \varepsilon(q)} q},
\label{plasmon_ch}
\end{equation}
where $m^{\ast}$ is the effective electron mass, $N$ is the equilibrium electron density in the 2DES. The effective dielectric function of the environment is
\begin{equation}
\varepsilon(q) = \frac{\varepsilon_{\rm s}+ \varepsilon \coth{(q h)}}{2},
\label{eps(q)}
\end{equation}
$\varepsilon_{\rm s}$ is the semiconductor substrate permittivity, $\varepsilon$ -- the dielectric permittivity of the spacer between the gate and the 2DES (for an ungated surface, $\varepsilon=1$), and $h$ the gate–2DES separation (see Fig.~\ref{Fig-grating}).

In the ungated regions, where the condition $q h \gg 1$ holds, the plasmon dispersion is of the standard square-root form~\cite{Stern1967, Chaplik1985}:
\begin{equation}
\omega = \sqrt{\frac{ 2\pi N_{\rm u} e^2 }{ m \varepsilon_{\rm eff} } q},
\label{plasmon}
\end{equation}
where $N_{\rm u}$ is the two-dimensional electron density in the ungated areas, and $\varepsilon_{\rm eff}=\left(\varepsilon_{\rm s}+ 1\right)/2$ is the effective dielectric permittivity. 

Under the gate, the opposite limit $q h \ll 1$ is typically realized. In this case the Coulomb interaction is screened by the nearby metal, and the system supports screened plasmons with a linear dispersion~\cite{Chaplik1972}:
\begin{equation}
\omega = s q, \qquad s= \sqrt{\frac{ 4\pi N_{\rm g} e^2 h}{m \varepsilon }},
\label{scr_plasmon}
\end{equation}
where $s$ is the plasmon phase velocity, $N_{\rm g}$ is the electron density under the gates, $h$ is the gate-to-channel separation. Equations~\eqref{plasmon} and \eqref{scr_plasmon} allow for different carrier densities in gated and ungated parts of the 2DES, $N_{\rm u} \neq N_{\rm g}$. Since $N_{\rm g}$ can be tuned by the gate bias applied to the metallic grating, the velocity $s$ becomes electrically controllable. This tunability provides a means to shift the allowed plasmon bands and band gaps of the plasmonic crystal. In particular, for a dual-grating structure, both gated-plasmon velocities, $s_1$ and $s_2$, can be controlled independently. 

The main approximation involved in applying the Kronig–Penney model to realistic structures (both gated–gated and gated–ungated) is the use of a simple ``matching'' of the alternating potential~\eqref{Eq_phi} and the plasma-wave current at the boundary between gated and ungated regions of the plasmonic crystal, rather than a rigorous solution of the full three-dimensional electrodynamic problem. Alternative analytical approaches to the boundary-matching problem have been considered in Refs.~\cite{Zabolotnykh2020, Semenenko2020, Rodionov2024, Moiseenko2025}. Notwithstanding this approximation, several recent studies by independent groups have confirmed the validity of the Kronig–Penney model for describing particular experiments~\cite{Dub2024, Khisameeva2025}.

\section{Dispersion equation of a plasmonic crystal}

To begin, we derive the dispersion relation for plasma waves in a plasmonic crystal whose unit cell comprises two distinct regions, ``1'' and ``2''. We consider the general case in which the local relation between the wave number $q$ and the gate-to-channel separation $h$ (and thus the plasmon dispersion) in each region is arbitrary (see also Refs.~\cite{Kachorovskii2012, Petrov2017, Gorbenko2024, Khisameeva2025}). 
At the interface between the two regions, the plasmon-induced {\textit{ac}} current must be continuous, and the oscillating electric potential must remain continuous. The corresponding boundary conditions can be written as
\begin{gather}
N_1 v_1 = N_2 v_2,
\label{Eq_J}
\
\varphi_1 = \varphi_2,
\end{gather}
where $N_{1,2}$ are the equilibrium 2DES densities, $v_{1,2}$ are the \textit{ac} carrier velocities at the boundary, and $\varphi_{1,2}$ are the \textit{ac} electric potentials evaluated at the same interface.

Next, we write the oscillating corrections to the concentration and velocity as
\begin{align}
&\delta N_n
=A_n e^{i q_n x}+B_n e^{-i q_n x},
\label{eq-n}
\\
&v_n=\frac{\omega}{q_n N_n}\left(A_n e^{i q_n x}- B_n e^{-i q_n x}\right),
\label{eq-v}
\end{align}
where $n=1,2.$ In these equations, we have taken into account that $\delta N$ and $v$ must satisfy the linearized continuity equation. To find the spectrum, we also need to determine the relationship between $\varphi_n$ and $\delta N_n.$ We assume (and this is essentially the only fundamental simplification of the Kronig-Penney model) that this relationship is the same as in the case of an infinite system: 
\begin{equation}
\varphi_n =K_n \delta N_n,
\qquad
K_n = \frac{2 \pi e^2 }{q_{n} \varepsilon(q_n)},
\label{Eq_phi}
\end{equation}
where $n=1,2,$ and the wave vectors $q_n$ are related to the frequency $\omega$ by the expression
\begin{equation}
\omega^2= \frac{K_n N_n q_n^2}{m},
\label{eq-omega}
\end{equation}
which holds for both $n=1$ and $n=2.$ In the gated-ungated case, we assume that $N_1=N_{\rm g}$, $h_1=h$, $q_1=q_{\rm g}$ and the spectrum is described by formula \eqref{scr_plasmon}, while in region ``2'' formula \eqref{plasmon} is reproduced with $N_2=N_{\rm u}$ and $q=q_{\rm u}$. In the case of a dual gate grating (gated-gated case), we have $N_1=N_{\rm g1},~N_2=N_{\rm g2},~h_1=h_2=h,$ and the wave vectors are $q_1=q_{\rm g1}$ and $q_2=q_{\rm g2}.$ In the latter case, the possibility arises to change the concentrations and plasma velocities in both region ``1'' and region ``2''.

Expressing $q_{ \rm u}$ and $q_{\rm g}$ from relations \eqref{plasmon} and \eqref{scr_plasmon}, we obtain:
\begin{align}
&q_{u} = \frac{m \varepsilon_{\rm eff}} {2 \pi N_{\rm u} e^2} \omega^2,
\label{q_ungated} 
\\
&q_{g} = \frac{\omega}{s}.
\label{q_gated}
\end{align}

The Bloch periodicity condition requires that, upon translation by one lattice period $L$, the amplitudes transform as
\begin{equation}
A_{n+1} = A_n e^{i q L}, \qquad B_{n+1} = B_n e^{i q L},
\end{equation}
where $q$ is the plasmon quasimomentum. For normal incidence of an electromagnetic wave on the crystal, the in-plane momentum transfer vanishes and thus $q=0$. From an experimental standpoint, a finite quasimomentum can be imparted to the crystal either by exploiting its finite size~\cite{Allen:1983}, or by imposing a controlled density modulation, via surface acoustic waves~\cite{Kukushkin}.

Standard transfer-matrix calculations then give the following expression for the plasmonic-crystal spectrum:
\begin{equation}
\cos{q_{1} L_{1}} \cos{q_{2} L_{2}} - Z \sin{q_{1} L_{1}} \sin{q_{2} L_{2}}
= \cos{q L},
\label{eq-spectr}
\end{equation}
where $Z$ is a dimensionless parameter that quantifies the coupling strength in the crystal ($Z \approx 1$ corresponds to weak coupling, whereas $Z \gg 1$ corresponds to strong coupling) and can be expressed in either of two equivalent forms:
\begin{equation}
Z= \frac{(K_1 q_1)^2 +(K_2 q_2)^2}{ 2 K_1 K_2 q_1 q_2 }= \frac{(N_1 q_1)^2 +(N_2 q_2)^2}{ 2 N_1 N_2 q_1 q_2 }.
\label{Z}
\end{equation}
Here we have used formula \eqref{eq-omega}, from which it follows that $ K_n q_n \propto 1/(N_n q_n).$ In the simplest case of a gated-gated structure, formula \eqref{Z} simplifies \cite{Kachorovskii2012}
\begin{equation}
Z=\frac{s_1^2 + s_2^2}{ 2 s_1 s_2}.
\end{equation}

\begin{figure*}
\centering
\hspace*{-0.1in}
\includegraphics[width=0.3\textwidth]{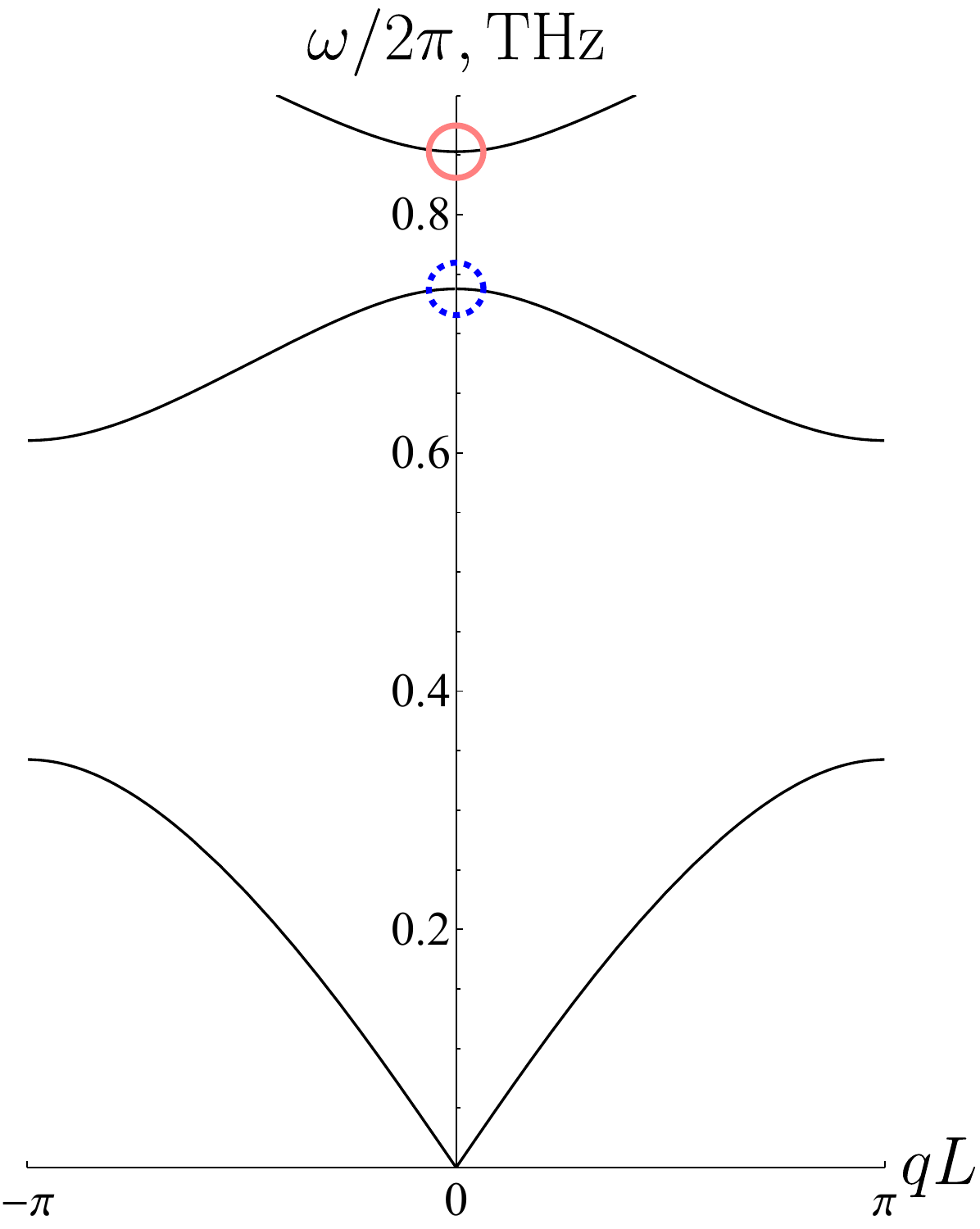}% Here is how to import EPS art
\includegraphics[width=0.3\textwidth]{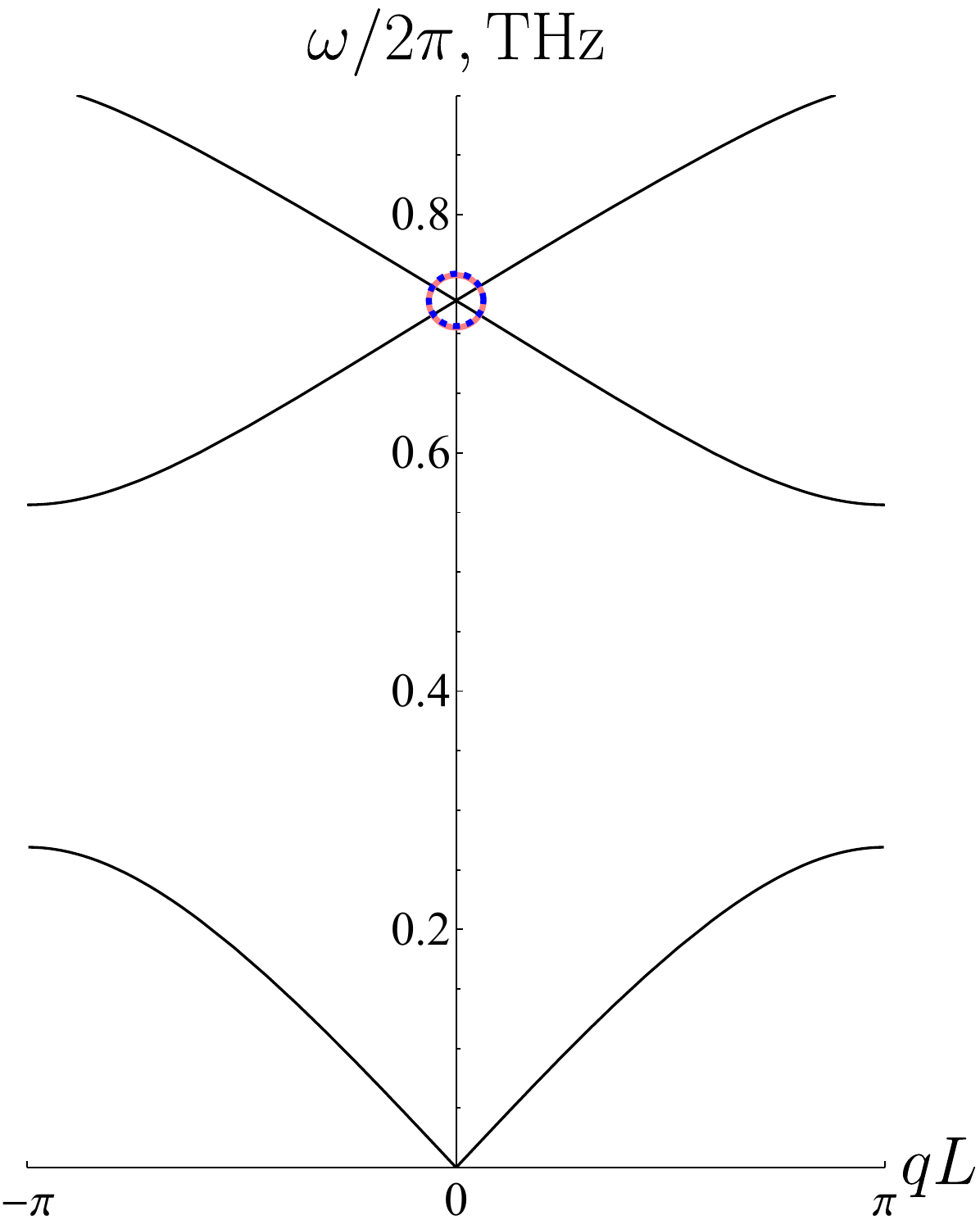}% Here is how to import EPS art
\includegraphics[width=0.3\textwidth]{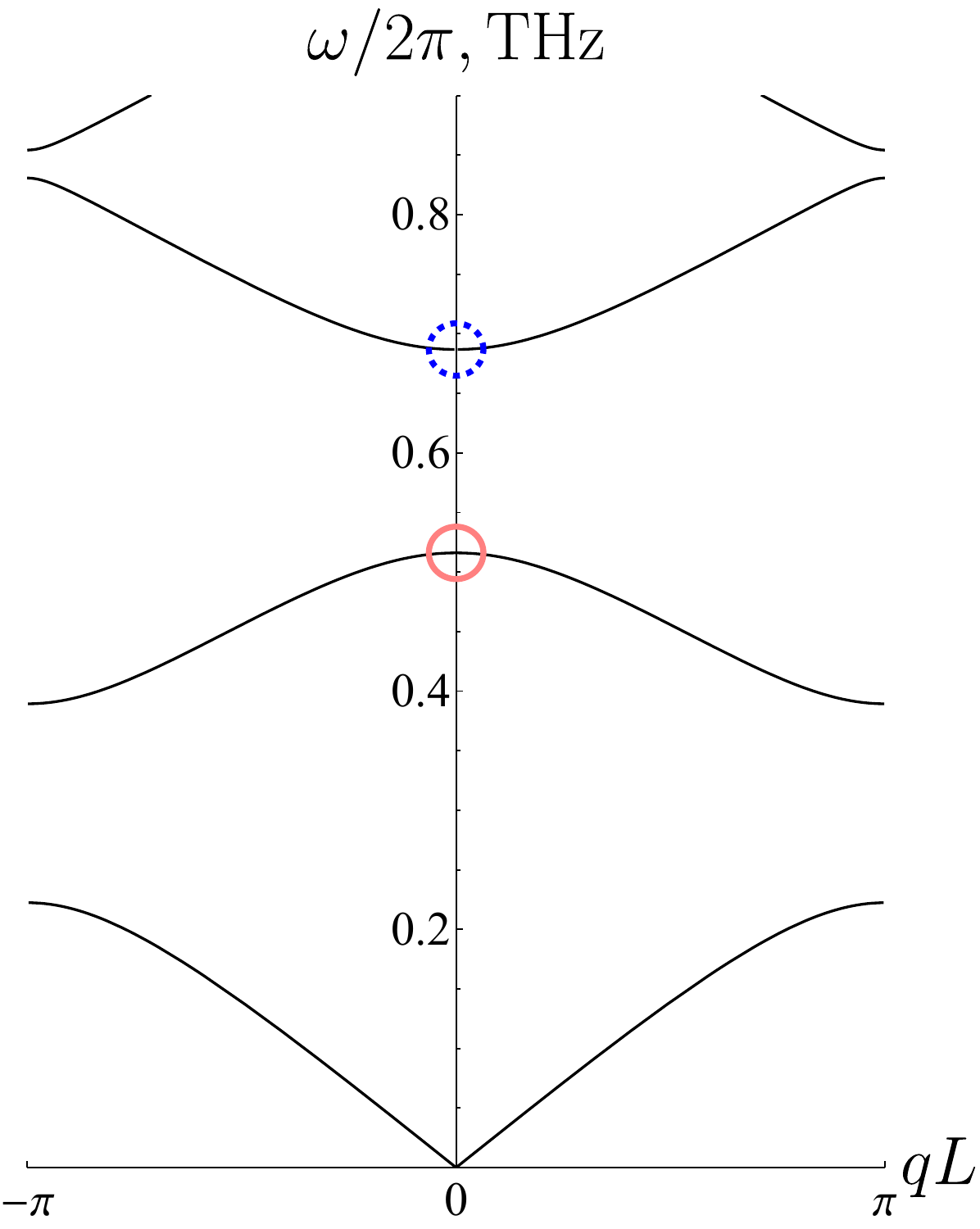}% Here is how to import EPS art
\caption{
Band diagrams of a gated-ungated plasmonic crystal, plotted using experimental parameters from Ref.~\cite{Khisameeva2025} for different filling factors. Left: $f=0.15$; center: $f=f^*=0.28$; right: $f=0.5$. Circles denote the dark and bright modes at $q = 0$ solid circle – bright mode, dashed circle – dark mode. The filling factor at which band crossing occurs, $f=f^*$ (central panel) corresponds to the condition $q_{\rm g} L_{\rm g} = q_{\rm u} L_{\rm u} = \pi$ according to Eq.~\eqref{Eq_case2} for $N = M =1$. Note that as the filling factor increases  above $f^*$, the dark state ``jumps'' from the lower branch to the upper one (a similar transition of the dark mode from one band to another can be observed by varying the parameter $\eta$, as shown in Fig.~10 of Ref.~\cite{Gorbenko2024}). Consistent with Fig.~\ref{Fig_3}, the effective mass of the bright plasmon is positive for $f<f^*$ and negative for $f>f^*$. For $f=f^*$ there are three linearly dispersing modes at small $q$: a low-frequency acoustic mode and two modes arising from the intersection of the bright and dark excitations.}
\label{Fig-dark_state_jump}
\end{figure*}

Weak coupling corresponds to $s_1\approx s_2,$, whereas strong coupling is achieved when the characteristic plasma wave velocities in the different parts of the unit cell differ substantially, for example, $s_1 \gg s_2$.

\subsection{Bright and dark plasmon modes}

The dispersion relation \eqref{eq-spectr} implicitly defines the plasmonic crystal spectrum $\omega(q)$. It is convenient to rewrite it in the form
\begin{equation}
 {\rm B}(\omega) {\rm D}(\omega) = \sin^2\left (\frac{q~L}{2}\right),
\label{Eq_BD}
\end{equation}
where
 \begin{equation}
B(\omega) = \cos{\frac{q_{1} L_{1}}{2}} \sin{\frac{q_{2} L_{2}}{2}} + \eta \sin{\frac{q_{1} L_{1}}{2}} \cos{\frac{q_{2} L_{2}}{2}}, 
\label{B}
\end{equation}
\begin{equation}
D(\omega) = \cos{\frac{q_{1} L_{1}}{2}} \sin{\frac{q_{2} L_{2}}{2}} + \frac{1}{\eta} \sin{\frac{q_{1} L_{1}}{2}} \cos{\frac{q_{2} L_{2}}{2}}.
\label{D}
\end{equation}
The dimensionless parameter
\begin{equation}
\eta= \frac{K_1 q_1}{K_2 q_2}= \frac{q_2 N_2}{q_1 N_1}
\label{eta}
\end{equation}
is related to the coupling parameter $Z$ introduced above via
\be
Z = \frac{1}{2} \left(\eta + \frac{1}{\eta} \right),
\label{Z-eta}
\ee 
where Eq.~\eqref{eq-omega} has been used in the last step.

For the gated–gated crystal configuration, the parameter $\eta$ takes a particularly simple form:
\begin{equation}
\eta=\frac{s_{2}}{s_{1}}.
\label{Eq_eta_gg}
\end{equation}
For the gated–ungated configuration, we set $q_{1}=q_{\rm g} $ and $q_2=q_{\rm u}$ and express these wave vectors in terms of $\omega$ using Eqs.~\eqref{plasmon} and \eqref{scr_plasmon}. This gives
\begin{equation}
\eta=\frac{2\omega h \varepsilon_{\rm eff}}{\varepsilon s_{1}} .
\label{Eq_eta_gu}
\end{equation}

Unlike the gated–gated case, the parameter $\eta$ depends on frequency in the gated–ungated structure.

The functions $B(\omega)$ and $D(\omega)$ have a clear physical meaning~\cite{Gorbenko2024}. Their zeros determine, respectively, the so‑called bright, $B(\omega_{\rm b})=0,$  and dark, ($D(\omega_{\rm d})=0,$ eigenmodes. For normally incident radiation in the long‑wavelength limit ($\lambda \gg L$), one may set $q=0$. In this case, Eq.~\eqref{Eq_BD} reduces to 
$B(\omega)D(\omega) =0$, yielding two series of modes. However, only the bright series is optically active~\cite{Gorbenko2024}. This selection rule is lifted in structures that lack inversion symmetry.

The parameter $\eta$ introduced above warrants additional attention. It can take any positive value $ 0<\eta<\infty $, and according to Eq. \eqref{Z-eta}, the transformation $\eta \rightarrow 1/\eta$ leaves the coupling parameter $Z$ unchanged. Thus, both limits $\eta \gg 1$ and $\eta \ll 1$ correspond to $Z \gg 1$, i.e.,the strong coupling regime. Consequently, the coupling regime is fully determined by the single parameter $Z$. At the same time, $\eta$ carries additional physical information. As we demonstrate in Section~\ref{sec-localization}, the regimes $\eta<1$ and $\eta>1$ (though they yield the same value of $Z$) correspond to plasmonic oscillations with qualitatively different spatial profiles within the unit cell of the crystal.

\begin{figure}[h!]
\centering
\includegraphics[width=6.6 cm]{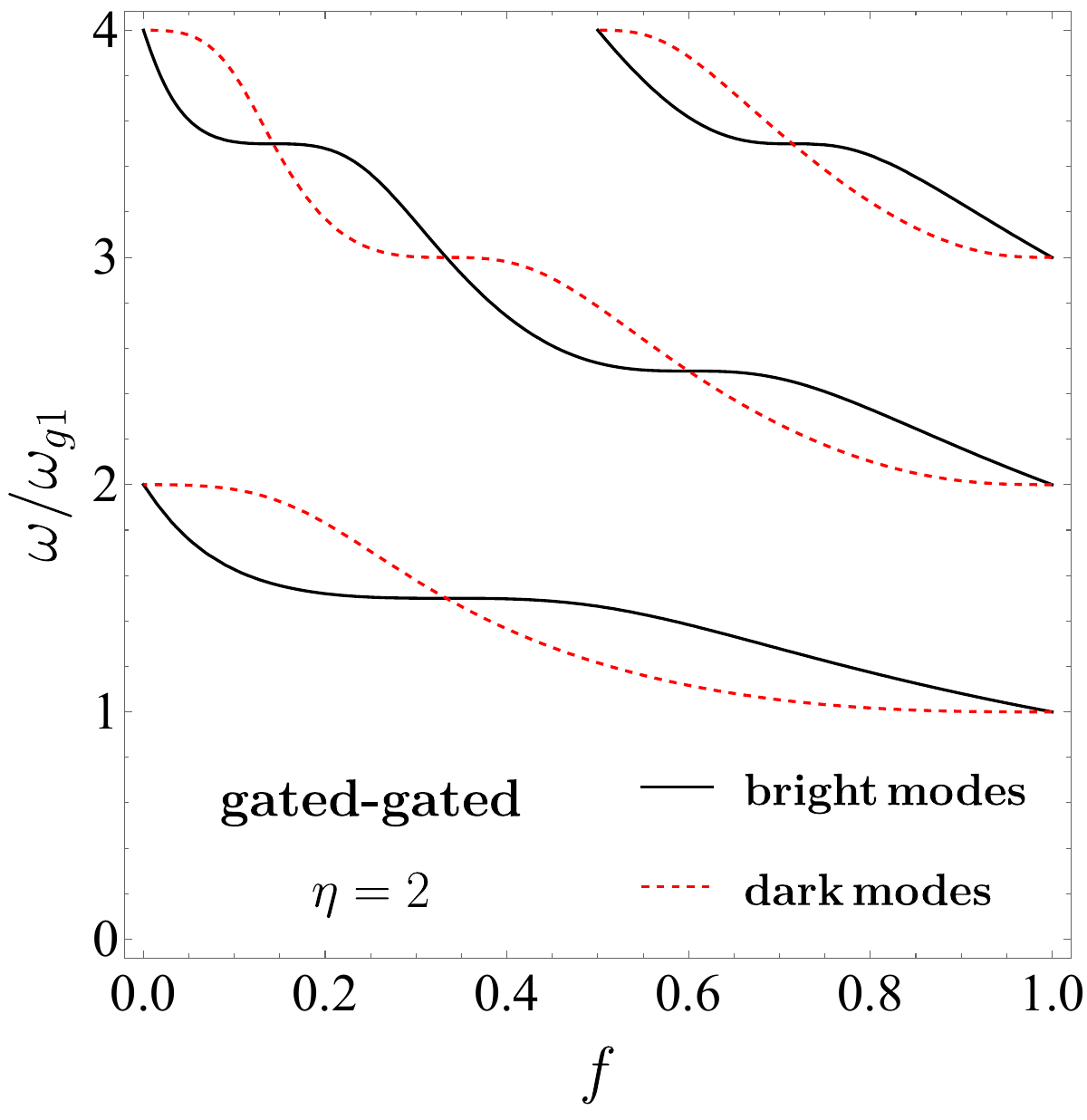}

\includegraphics[width=6.6 cm]{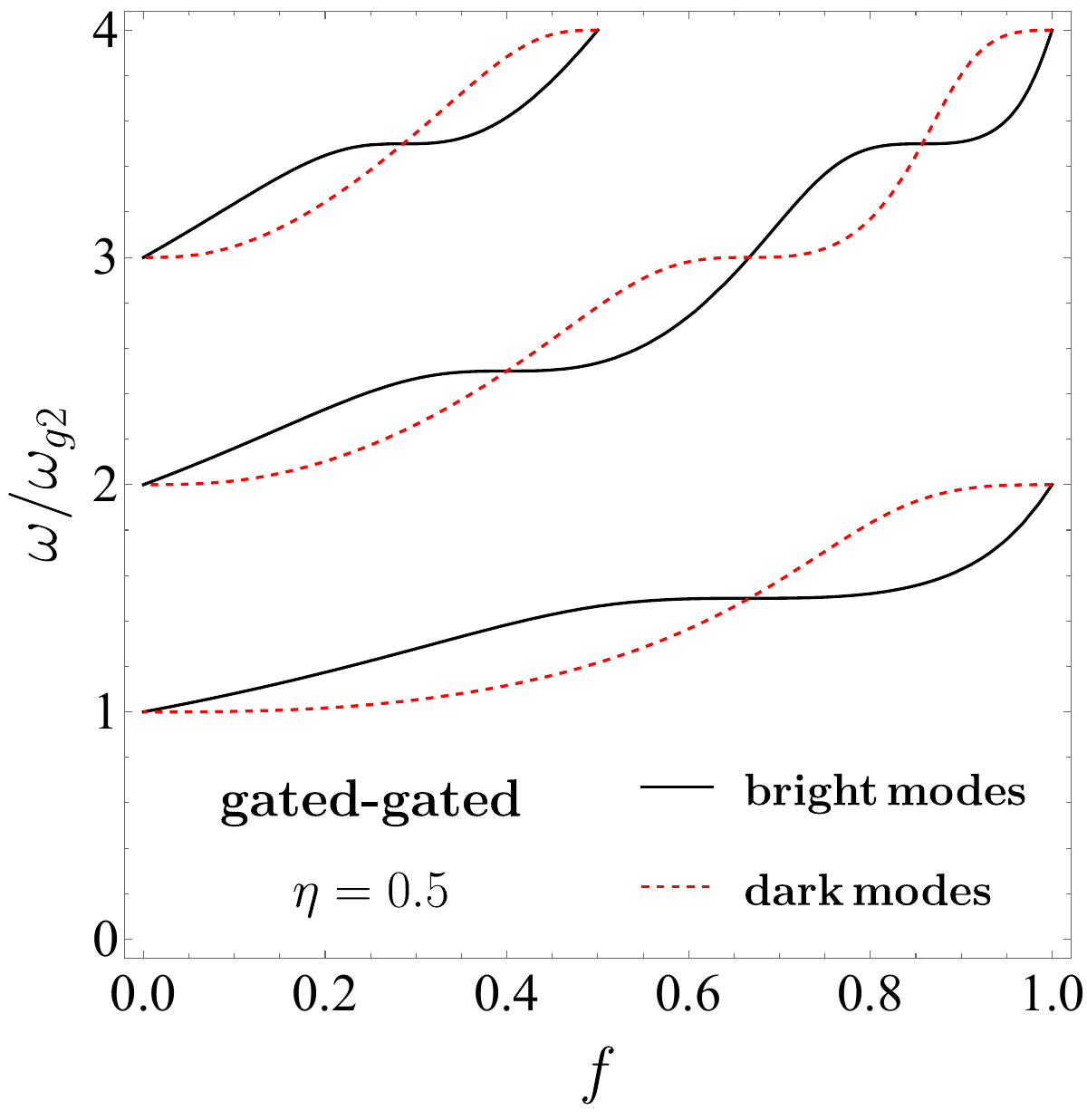}% Here is how to import EPS art
\caption{Dependence of the frequencies of the several lowest plasma excitations at $q=0$ on the filling factor for the gated-gated case: top, $s_1 = 0.5 s_2$; bottom, $s_1 = 2 s_2$.}
\label{Fig-w(f)_gg}
\end{figure}

\begin{figure}[h!]
\centering
\includegraphics[width=6.6 cm]{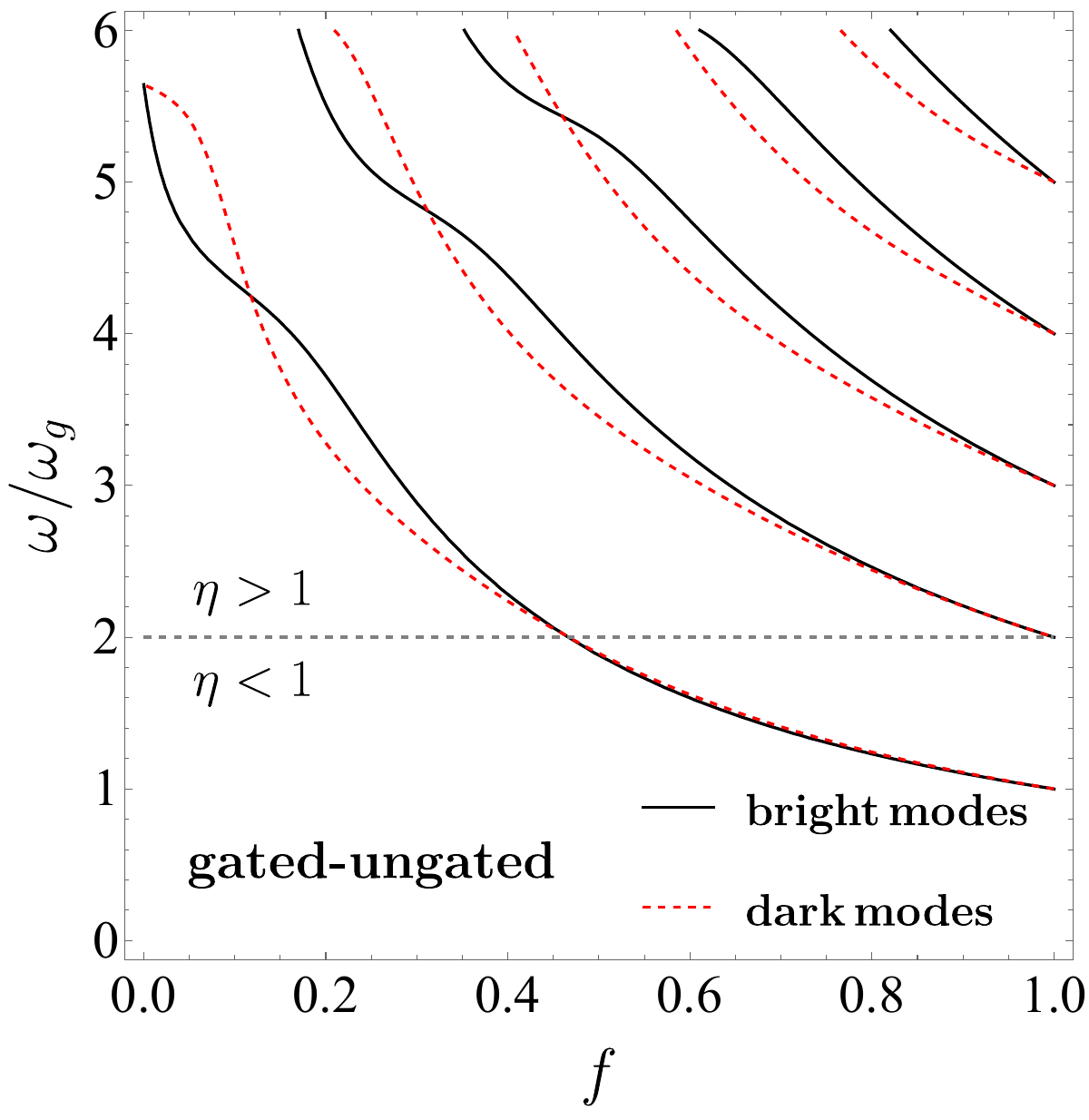}
\includegraphics[width=6.6 cm]{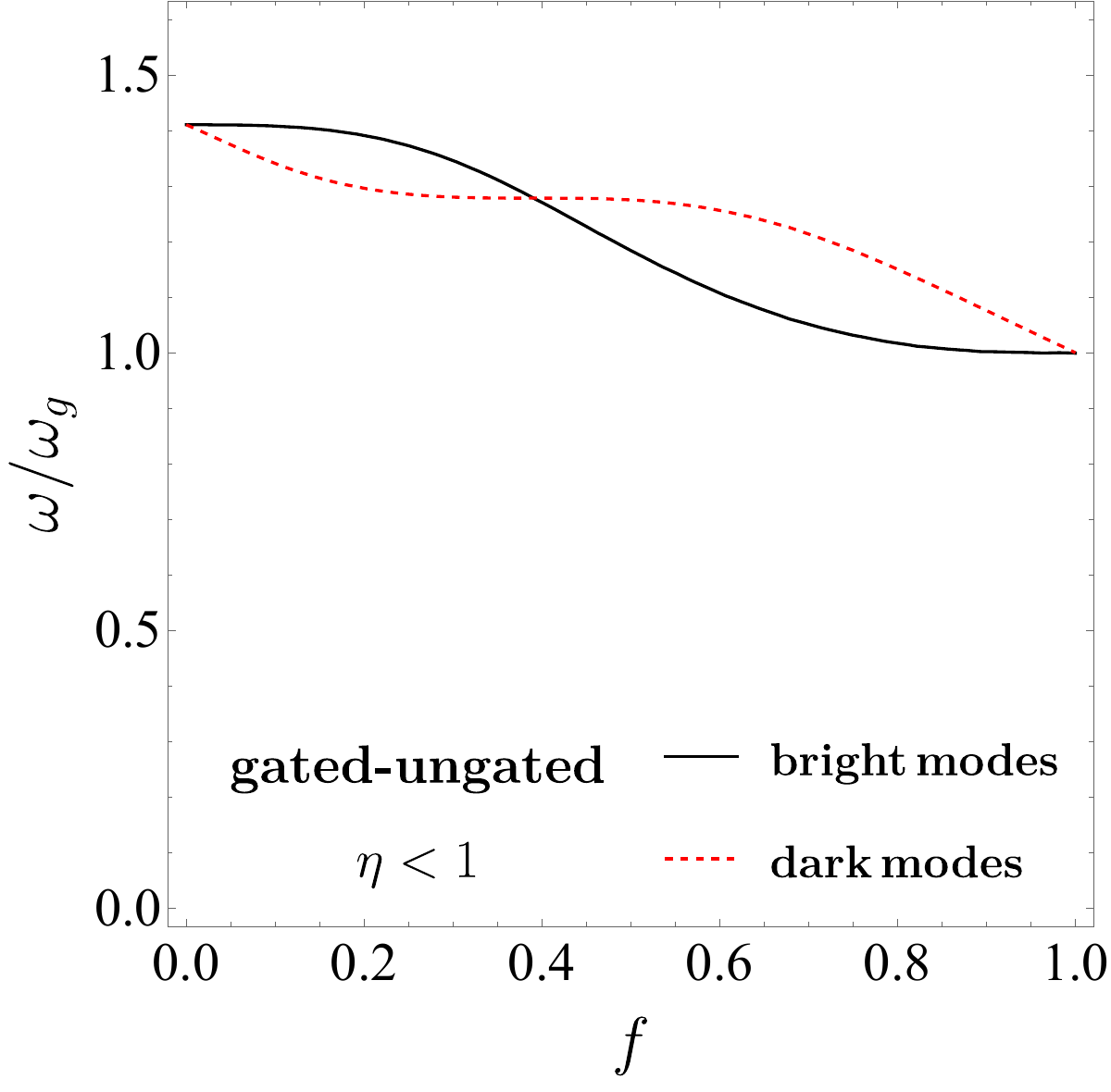}
\caption{Dependence of the plasmonic crystal eigen-frequencies at  $q=0$ on the filling factor for the gated-ungated configuration ($h = L/25, \varepsilon^s = \varepsilon$): top, $N_{\rm u}/N_g =16$; bottom, $N_g = N_{\rm u}$. In the top panel, the horizontal dotted line divides the figure into two regions: below the line, $\eta <1$; above it, $\eta>1$  (in the gated-ungated case, $\eta$ depends on frequency according to Eq. \eqref{Eq_eta_gu}). In the bottom panel, region $\eta>1$  corresponds to  very high frequencies: $\omega \gtrsim 2\omega_g$.}
\label{Fig-w(f)_gu}
\end{figure}

\subsection{Fundamental acoustic plasmon mode}

In the low-frequency limit, for both gated–gated and gated–ungated configurations, the product $B(\omega) D(\omega)$ scales as $\omega^2$. Therefore, the lowest plasmon branch exhibits a linear (acoustic) dispersion at small quasimomentum $q$ (see the low-frequency region in Fig.\ref{Fig-dark_state_jump}):
\begin{equation}
\omega = s_{\rm ac} q.
\end{equation}
The acoustic-mode velocity follows directly from Eq.~\eqref{Eq_BD} as
\begin{equation}
s_{\rm ac} = \frac{L}{2 \sqrt{ B'(0) D'(0)}}.
\label{Eq_s_ac}
\end{equation}
For the gated–gated configuration, this formula yields
\begin{equation}
\begin{aligned}
&s_{\rm ac} = \sqrt{\frac{1}{{f}/{s_1^2}+{(1-f)}/{s_2^2}}}
\\
&= s_{1} \sqrt{\frac{1}{{f}+{(1-f)}(N_{1}/N_{2})}},
\end{aligned}
\label{eq:s-ac-gg}
\end{equation}
whereas for the gated–ungated configuration we obtain
\begin{equation}
s_{\rm ac} = s_g \sqrt{\frac{1/f}{ f+ (1-f)(N_{\rm g}/N_{\rm u})}}.
\label{eq:s-ac-gu}
\end{equation}
As shown below, for certain gate voltage values, the system can support not only the fundamental acoustic plasmonl, but also higher-order modes with an approximately linear dispersion (see the central panel in Fig.~\ref{Fig-dark_state_jump}).

\section{Effective Plasmon Mass}
\subsection{Effective mass of bright and bark plasmon modes}

The plasma wave spectrum, $f(q)=\omega(q)/(2\pi)$, determined by Eq.~\eqref{Eq_BD}, is plotted in Fig.~\ref{Fig-dark_state_jump} for the gated-ungated geometry. The calculations use the experimental parameters from Ref.~\cite{Khisameeva2025} for several filling factors. The solid pink and dashed black circles mark the $q=0$ eigenfrequencies of the lowest bright and bright plasmon modes, denoted below  $\omega_{\rm b}$ and $\omega_{\rm d}$, respectively.

We now derive the effective masses of the bright and dark plasmon modes. For this purpose, we expand the dispersion equation \eqref{Eq_BD} near $q=0$ and near each frequency root. Since the bright  and dark modes at $q=0$ are defined by $B(\omega_{\rm b})=0$ and $D(\omega_{\rm d})=0$, respectively, the leading-order expansion for small $q$ gives $(\omega-\omega_{\rm b}) {\rm B}'(\omega_{\rm b}) {\rm D}(\omega_{\rm b})= {q^2 L^2}/{4}$ for the bright mode and   $ (\omega-\omega_{\rm d}) {\rm B}(\omega_{\rm d}) {\rm D}'(\omega_{\rm d})= {q^2 L^2}/{4}$ for the dark one. These equations directly yield
 the effective masses of the bright and bright plasmons:
\begin{equation}
\begin{aligned}
&m_{\rm b} = \frac{2 \hbar B'(\omega_{\rm b}) D(\omega_{\rm b})}{L^2},
\\
&m_{\rm d} = \frac{2 \hbar B(\omega_{\rm d}) D'(\omega_{\rm d})}{L^2}.
\end{aligned}
\label{Eq_ms}
\end{equation}
The frequencies $\omega_{\rm b}$ and $\omega_{\rm d}$ depend on $L_{1,2}$ and therefore on the filling factor $f=L_1/L$, as well as on the parameter $\eta$. The resulting dependence of the effective masses of $m_{\rm b}$ and $m_{\rm d}$ on $f$ (with other parameters fixed) is shown in Figs.~\ref{Fig-m(f)_gg} and \ref{Fig-m(f)_gu} for the gated-gated and gated-ungated geometries, respectively.

Analytical estimates for the effective masses in the weak- and strong-coupling regimes are derived for the gated-gated case in Section~\ref{sec-gated}. For the gated-ungated structure, corresponding estimates are provided in Section~\ref{app:experiment}, with particular emphasis on the recent experiment reported in Ref.~\cite{Khisameeva2025}.

\subsection{Degeneracy Points. Modes with Linear Dispersion.}
\label{sec:deganeracy_points}
\begin{figure}[h!]
\centering
\includegraphics[width=6.6 cm]{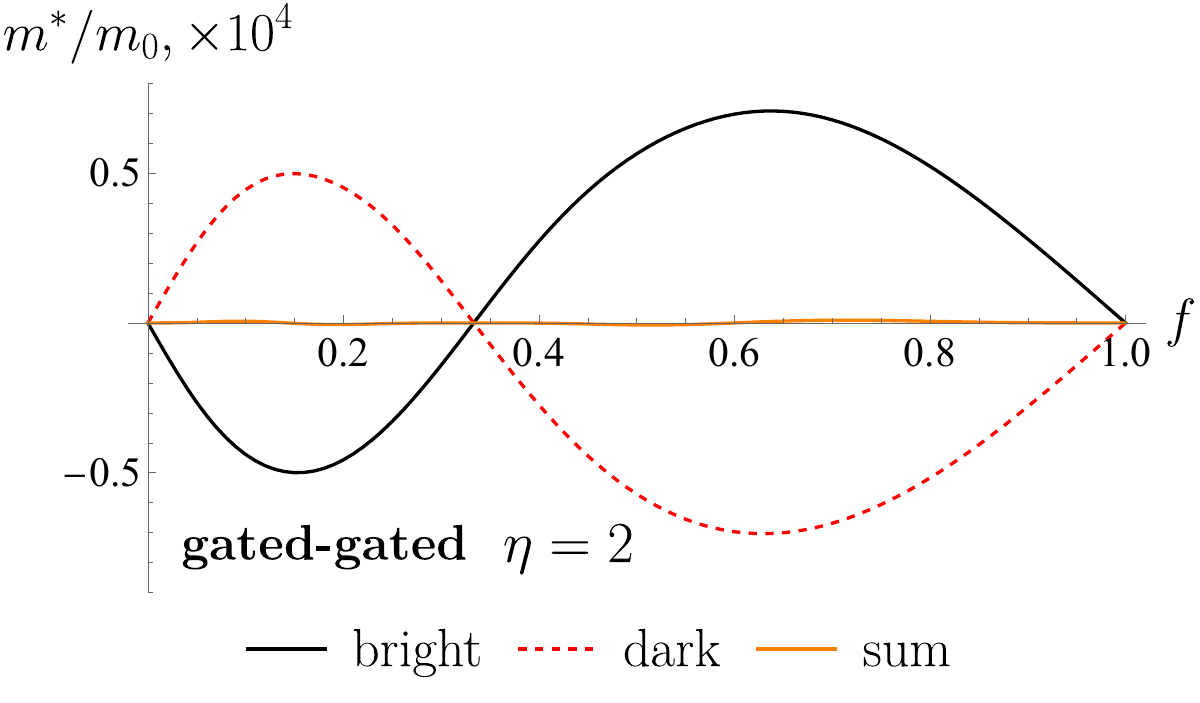}

\includegraphics[width=6.6 cm]{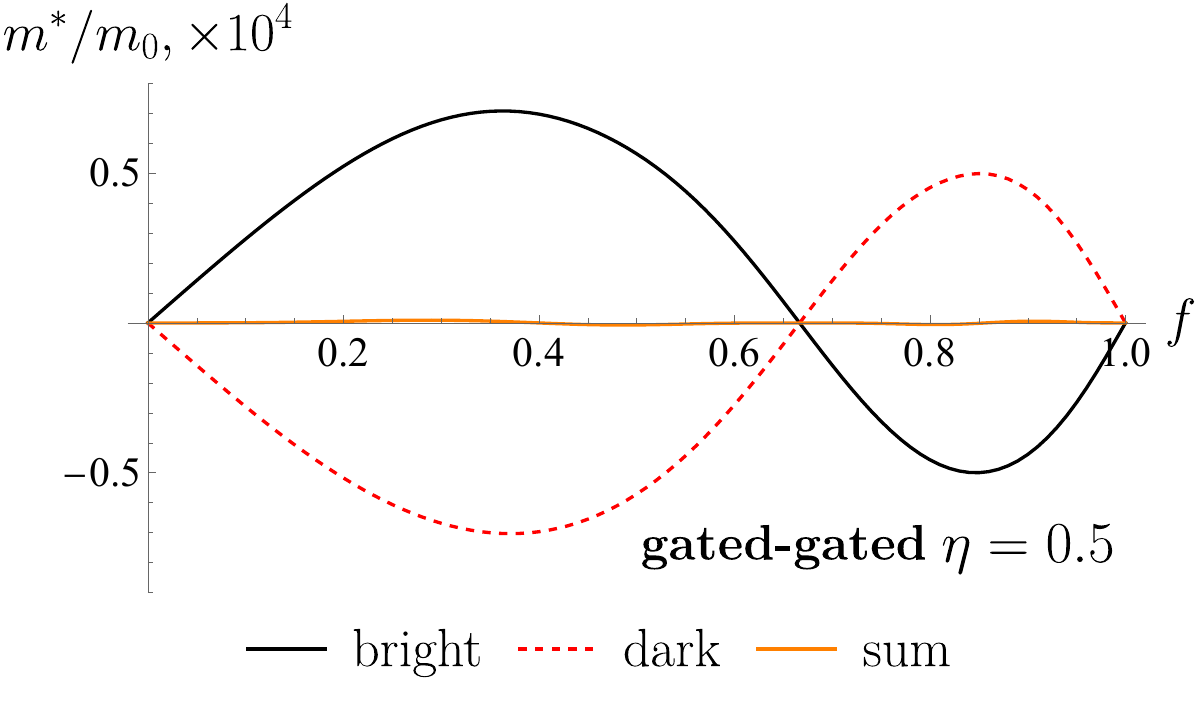}% Here is how to import EPS art
\caption{Dependence of plasmon effective masses, normalized on $m_0$ - the effective electron mass in GaAs, on the filling factor for the gated-gated case: top, $s_1 = 0.5 s_2$; bottom, $s_1 = 2 s_2$.}
\label{Fig-m(f)_gg}
\end{figure}

\begin{figure}[h!]
\centering
\includegraphics[width=8.2 cm]{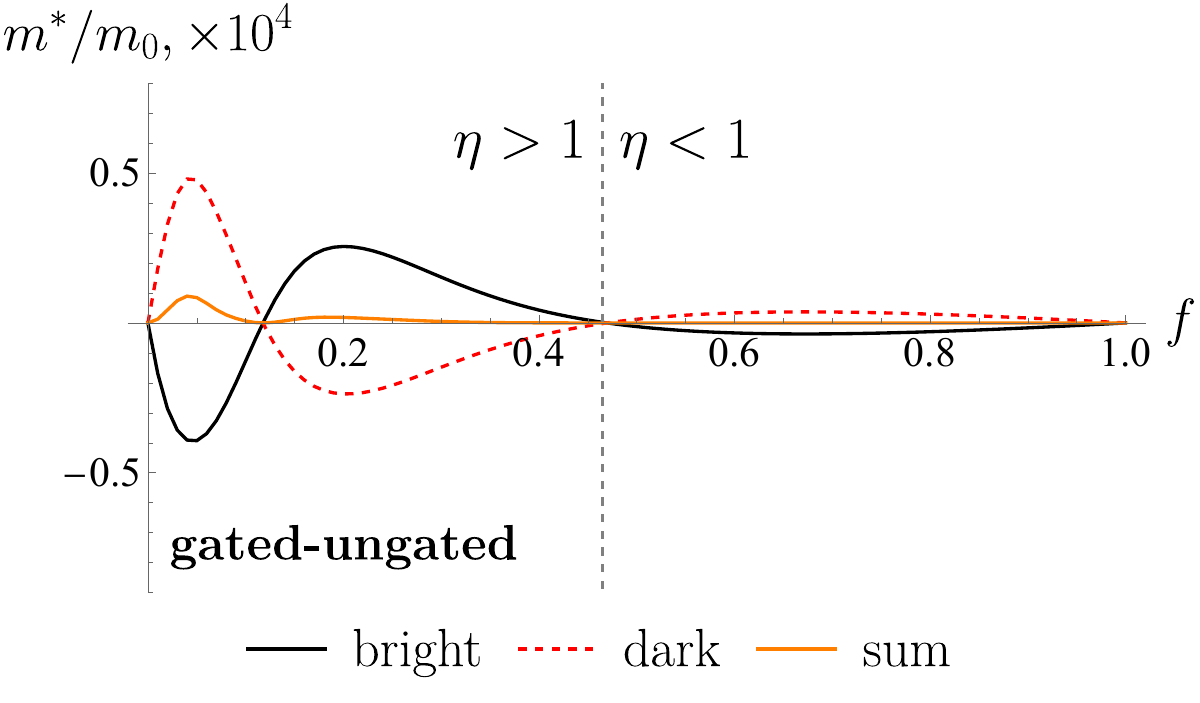}

\includegraphics[width=8.2 cm]{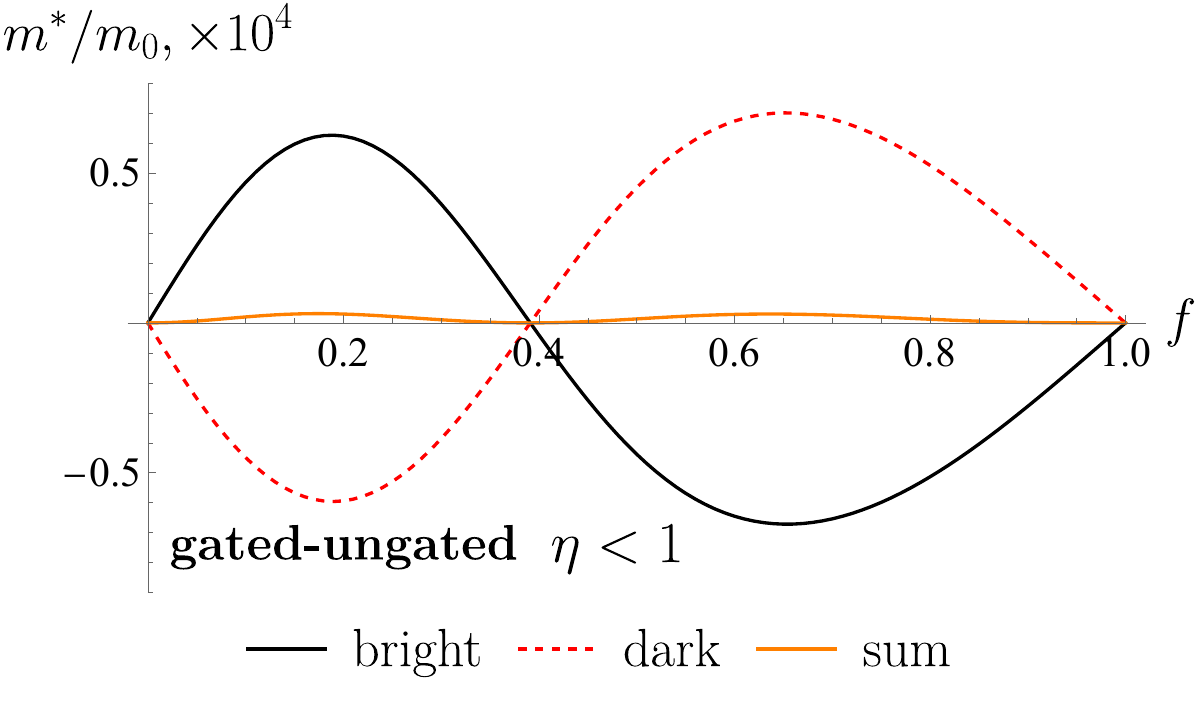}% Here is how to import EPS art
\caption{Dependence of plasmon effective masses, normalized on $m_0$ - the effective electron mass in GaAs, on the filling factor for the gated-ungated case ($h = L/25, \varepsilon^s = \varepsilon$): top, $N_{\rm u}/N_g =16$; bottom, $N_g = N_{\rm u}$.}
\label{Fig-m(f)_gu}
\end{figure}

As shown in Figs.~\ref{Fig-m(f)_gg} and \ref{Fig-m(f)_gu}, the frequencies intersect at specific values of $f$, yielding $\omega_{\rm b}=\omega_{\rm d}=\omega_0.$ There are three intersection points: two trivial points corresponding to the limiting filling factors  $f=0,$ and  $f=1,$ and less trivial point $f=f^*,$ which was already discussed above (see central panel in Fig.~\ref{Fig-dark_state_jump}).     At all these  points, the plasmon mass vanishes: $m_{\rm b} (\omega_0) =m_{\rm d}(\omega_0)=0.$
We notice that the frequency  $\omega_0$  differs for each of these three points (see Eqs.~\eqref{Eq-small-f}, \eqref{Eq-large-f} and \eqref{Eq_w0_2} below).

Expanding Eq. \eqref{Eq_BD} to second order in $\omega-\omega_0$ and $q$ gives
$
(\omega-\omega_0)^2 {\rm D}'(\omega_0) {\rm B}'(\omega_0)= {q^2 L^2}/{4}.
$
Consequently, at this point the spectrum is linear and characterized by a velocity $s_0$, analogous to $s_{\rm ac}$ in Eq.\eqref{Eq_s_ac}:
\begin{equation}
|\omega-\omega_0|= s_0 q, \quad s_0 = \frac{L}{2 \sqrt{B'(\omega_0)D'(\omega_0)}}.
\end{equation}
Although the spectrum is linear near $\omega_0$, it is important to note that this mode is not acoustic (unlike the fundamental acoustic mode) due to its non-zero frequency $\omega_0$.

Let us discuss the most interesting case of filling factor $f=f^*.$  In the vicinity of this point, $f=f^*+ \delta f$, corrections to the mode frequencies arise: $\omega_{\rm b,d}=\omega_0 + \delta \omega_{\rm b,d}$. This leads to a small effective mass, which can be calculated from the equation
$
(\omega-\omega_0-\delta \omega_{\rm b} )(\omega-\omega_0-\delta \omega_{\rm d}) {\rm D}'(\omega_0) {\rm B}'(\omega_0) \approx {q^2 L^2}/{4}.
$
Solving the quadratic equation for $\omega-\omega_0$ and expanding the roots in $q$, we obtain
\begin{equation}
\begin{aligned}
& m_{\rm b,d} =\pm (\delta \omega_{\rm b} - \delta \omega_{\rm d}) \frac{2\hbar {\rm D}'(\omega_0) {\rm B}'(\omega_0)}{ L^2}=
\\
&
\pm \delta f \left [\frac{\partial (\omega_{\rm b} - \omega_{\rm d})}{\partial f} \right]_{f=f_0,\omega=\omega_0} \mkern-60mu \times \frac{2\hbar {\rm D}'(\omega_0) {\rm B}'(\omega_0)}{ L^2}
\end{aligned}
\label{Eq_m_eff_general}
\end{equation}
Thus, near the degeneracy point, the effective masses are equal in magnitude but opposite in sign.

In general, Eq.~\eqref{Eq_BD} must be solved numerically. However, frequencies of degeneracy of dark and bright modes can be found analytically. A detailed analysis of the different degeneracy cases and the corresponding solutions is provided in Appendix \ref{sec_modes_degeneracy}. 

Here we limit ourselves to the most interesting case: the degeneracy of the fundamental bright and dark modes. The corresponding degeneracy condition is
\be
q_1 L_1=q_2 L_2=\pi.
\ee
 This equation determines degenerate frequency $\omega_0$, which also satisfies $B(\omega_0)=D(\omega_0)=0$.
Near this frequency, when $\omega_0=\omega_{\rm b} = \omega_{\rm d},$ the spectrum is linear and characterized by a velocity $s_0$. This velocity differs for gated-gated and gated-ungated structures:
\be
\begin{aligned}
& s_0 = s_g \left( \frac{\varepsilon}{8 \pi \varepsilon_{\rm eff}} \frac{N_{\rm u}^3}{N_{\rm g}^3} \frac{L}{h} \right)^{1/4}, \quad \text{(gated-ungated)},
\\
&s_0 = \sqrt{s_1 s_2}, \quad \text{(gated-gated)}.
\end{aligned}
\label{eq-s0}
\ee

\section{Effective Plasmon Mass in Gated-Gated Structure}
\label{sec-gated}

\subsection{Weak coupling regime}
The weak coupling regime is realized when $\eta \approx 1 .$ In this case, plasma waves are weakly scattered at the boundaries between regions ``1'' and ``2''. In the limit $\eta \to 1,$ we have
\be
B(\omega) =D(\omega) = \sin \left(\frac{q_1 L_1 +q_2 L_2 }{2}\right). \ee
Therefore, the frequencies of the dark and bright modes are degenerate and can be found from the condition 
\be
q_1 L_1 +q_2 L_2 = 2\pi n.
\label{eq_cond}
\ee

Consider a gated-gated structure and assume
\be
s_1= s_0 + \delta s/2, \quad s_2= s_0 - \delta s/2, \qquad \delta s \ll s_0.
\ee
For $\delta s \to 0$, Eq.~\eqref{eq_cond} yields
\be
\omega_0= \frac{2 \pi N s_0 }{L},
\ee
where $N$ is an integer.
For small but finite $\delta s,$ we seek a solution to Eq.~\eqref{Eq_BD} in the form $\omega =\omega_0 + \delta \omega. $ Expanding this equation to second order in $\delta \omega,$ $\delta s$ and $q,$ and after simple transformations, we obtain the massive plasmon spectrum in the form
\be
\hbar \left[\delta \omega - \frac{\pi \delta s (2 f -1)}{L} N \right] = \pm s_0 \sqrt{\hbar^2 q^2 + (M s_0)^2 }
\label{eq_dirac}
\ee
where 
\be
M=\frac{\hbar~ \delta s }{s_0^2 L} \sin\left[ \pi N (2f -1)\right],
\ee
so that the masses of the dark and bright modes are given by $\pm M.$

\subsection{Strong coupling regime.}
This regime is realized when either $\eta \gg 1$ or $\eta \ll 1$. As we will see, it is sufficient to consider only one of these cases, for example, $\eta \ll 1$.

In the case $s_2 \ll s_1$, the equations
$B(\omega)=0$ and $D(\omega)=0$ simplify (see \eqref{B} and \eqref{D}) and become, respectively:
$\cos(q_1 L_1/2 ) \sin(q_2 L_2/2)=0$ and $\cos(q_2 L_2/2 ) \sin(q_1 L_1/2)=0.$
Thus, we find two series of bright solutions: $\omega_{\rm b 1}= \pi s_1( 2 N-1)/L_1,~\omega_{\rm b2}=2\pi s_2 M/L_2 $ and two series of dark solutions:
$\omega_{\rm d1}=2\pi s_1 M/L_1,~\omega_{\rm d 2}= \pi s_2( 2 N-1)/L_2, $ where $N,M=1,2,\dots$ are integers. Using Eqs.~\eqref{Eq_ms}, we obtain approximate expressions for masses:
\begin{align}
&m_{\rm b1}= -\frac{\hbar (2-f)}{2 L s_1 \eta} \sin \left [\frac{\pi (1-f) ( 2N-1) }{f \eta}\right],
\\
&m_{\rm b2}= \frac{\hbar (1-f)}{2 L s_1 \eta^2} \sin\left[\frac{2\pi f \eta M }{1-f } \right] \approx \frac{ \pi \hbar f M }{ L s_2 },
\\
&m_{\rm d1}= \frac{\hbar (2-f)}{2 L \eta s_1} \sin \left[\frac{2\pi M (1-f)}{\eta f}\right] ,
\\
&m_{\rm d2}=-\frac{\hbar (1-f)}{2L s_1 \eta^2} \sin\left[
\frac{\pi (2 N-1) \eta f }{1-f}
\right] \approx
-\frac{\hbar \pi ( 2 N-1) f }{2L s_2 } ,
\end{align}
Consequently, the masses of the fundamental bright and dark modes ($N=M=1$) are
$m_{\rm b}^*=-m_{\rm d}^* \approx { \pi \hbar f }/{( L s_2) }.$ Note that
\be
\frac{m_{\rm b1}}{m_{\rm b2}}
\sim \frac{m_{\rm d1}}{m_{\rm d2}} \sim \frac{\eta f}{1-f} \ll 1.
\ee

The case $\eta \gg 1$ is described by the same formulas, with the replacements: $s_1 \to s_2,~ s_2 \to s_1,~ \eta \to 1/\eta$ and $f\to 1-f.$

\section{Effective Plasmon Mass in Gated-Ungated Structure}
\label{app:experiment}
In this section, we focus on calculating the effective mass of the fundamental dark and bright modes in the gated-ungated structure.
In such a structure, the parameter $\eta$ depends on frequency according to Eq.~\eqref{Eq_eta_gu}. This allows a system with fixed parameters $N_{1,2}, L_{1,2}$ to be in intermediate coupling regime, $\eta \sim 1,$ at frequencies $\omega \sim \omega_0$, and in a strong coupling regime both at low frequencies ($\omega \ll \omega_0,$ when $\eta \ll 1$) and at high frequencies ($\omega \gg \omega_0,$ when $\eta \gg 1$). Here $\omega_0$ is determined from the condition $\eta =1$:
\be
\omega_0 = \frac{s_1}{2 h} \frac{\varepsilon_{\rm eff}}{\varepsilon}.
\label{Eq_w0_1}
\ee
The weak coupling regime is quite difficult to realize in such a structure, since the condition $\eta \approx 1$ is satisfied only in a narrow frequency interval $|\omega-\omega_0| \ll \omega_0$, which implies fine-tuning of the frequency or gate voltage.
For definiteness, we consider a realistic experimental situation. Specifically, we provide estimates for the conditions of Ref.~\cite{Khisameeva2025}, where the condition $N_{\rm u} = N_g$ was met, and the parameter $\eta$ was less than unity, $ \eta \lesssim 1.$
Consequently, these experimental conditions represent an intermediate coupling regime, which lies outside the scope of the analytical formulas derived in the previous section for the strong and weak coupling regimes. Nevertheless, asymptotic formulas can be obtained for various filling factors.

The general formula \eqref{Eq_ms} allows for the numerical computation of the plasmon effective mass for an arbitrary filling factor $0<f<1$. In contrast, the approximate expression \eqref{Eq_m_eff_general} yields simple analytical formulas near three degeneracy points: $f=0$, $f=1$, and $q_{\rm u} L_{\rm u} = q_{\rm g} L_{\rm g} = \pi$ (the case $N=M=1$ in Eq.~\eqref{Eq_case2}).
Below, we consider different limiting cases. For convenience, we introduce the frequencies
\begin{equation}
\begin{aligned}
&\omega_{\rm u} = \omega_{\rm u}(q)|_{q=2 \pi/L}, \\
&\omega_{\rm g}=\omega_{\rm g}(q)|_{q=2 \pi/L},
\label{Eq_wg_wu}
\end{aligned}
\end{equation}
where $\omega_{\rm u}(q)$ and $\omega_{\rm g}(q)$ are given by Eqs.~\eqref{plasmon}, \eqref{scr_plasmon}.

%\noindent
\underline{\textit{Small filling factor}, $f \ll 1$}.
For $f \ll 1,$ each of the frequency solutions given with Eq.\ref{Eq_w_u} splits into dark and bright frequencies. For the fundamental modes, we get:
\be
\omega_{\rm b} = \omega_{\rm u},\quad \omega_{\rm d} = \omega_{\rm u}-\frac{\omega_{\rm u}}{2}\left(\frac{\omega_{\rm u}^2}{\omega_g^2} -1 \right) f,
\label{Eq-small-f}
\ee
where the parameter ${\omega_{\rm u}}/{\omega_g}$ is greater than unity for practically all realistic experimental parameters when $L \gg h$:
\be
\frac{\omega_{\rm u}}{\omega_g} = \frac{L \varepsilon}{4 \pi h \varepsilon_{\rm eff}} > 1.
\ee
In particular, for experimental values $L = 8~ \mu$m, $h=220~$nm, $\varepsilon =\varepsilon_{\rm eff},$ we get $\omega_{\rm u}/\omega_g \approx 3$.

\begin{figure}[!t]
\center
\includegraphics[width=84mm]{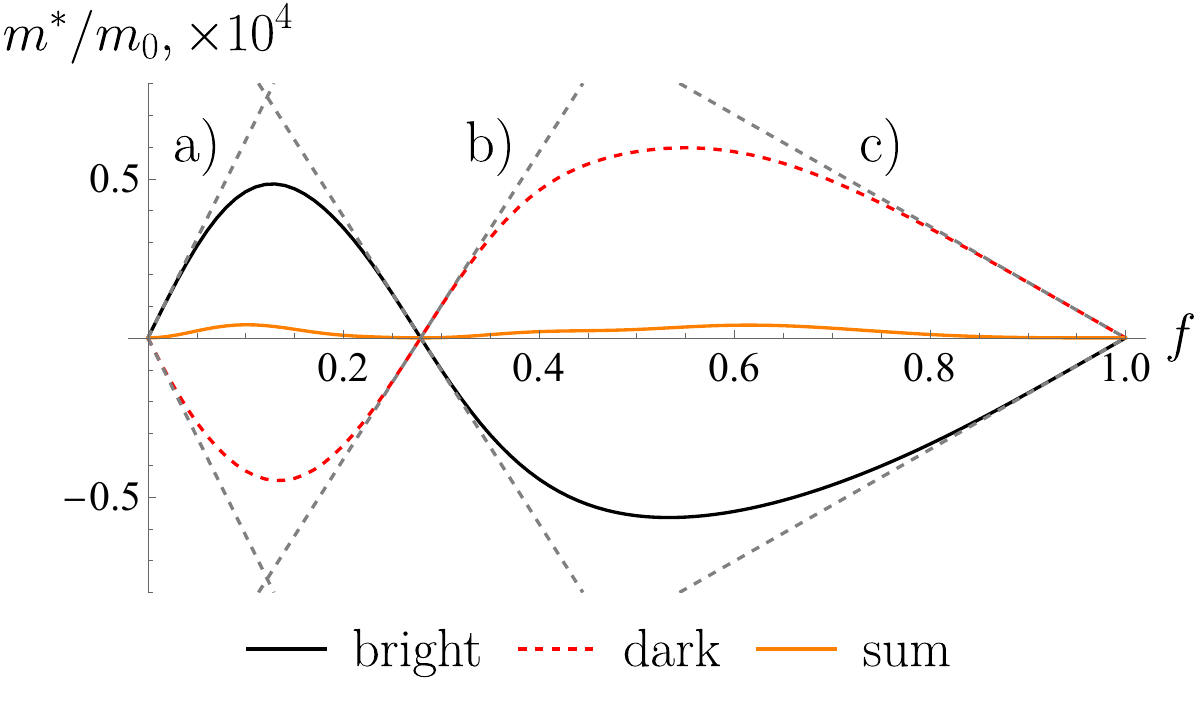}
\caption{The dependence of the ratio of the effective masses of the dark and bright plasmons to the effective electron mass in GaAs $m_0$ on the filling factor $f$ for the gated-ungated case. The asymptotic limits correspond to: a) $f\approx 0$: $|m_{\rm b,d}| = \frac{\hbar f}{\omega_{\rm u}} \left(\frac{2 \pi}{L}\right)^2 \left(\frac{\omega_{\rm u}^2}{\omega_g^2} -1\right)$; b) $q_{\rm g} L_{\rm g} \approx q_{\rm u} L_{\rm u}\approx \pi$: $|m_{\rm b,d}| = \frac{\hbar (f-f^*)}{ \omega_g} \left(\frac{2 \pi}{L}\right)^2 \frac{(2-f^*)(1-2 f^*)}{4 (1-f^*)^2 f^* } $, where $f^*$ is determined by Eq.~\eqref{Eq_fs}; c) $f\approx 1$: $|m_{\rm b,d}| = \frac{\hbar (f-1)}{2 \omega_g} \left(\frac{2 \pi}{L}\right)^2 \left(\frac{\omega_{\rm u}^4}{\omega_g^4} -1\right)$.}
\label{Fig_3}
\end{figure}

When $\omega_{\rm b} \neq \omega_{\rm d}$, the spectrum $\omega(q)$ ceases to be linear and becomes parabolic, which allows for the computation of non-zero effective masses. To linear order in the parameter $f$, these masses have opposite signs and the same absolute value:
\be
m_{\rm b} = - m_{\rm d} = \frac{f}{ \omega_{\rm u}} \left( \frac{2 \pi}{L}\right)^2 \left(\frac{\omega_{\rm u}^2}{\omega_g^2} -1 \right) >0
\ee
In the quadratic order in $f \ll 1$, sum of two masses becomes non-zero:
\be
\frac{m_{\rm b} + m_{\rm d}}{m_{\rm b}} = \frac{f}{2} \left(\frac{\omega_{\rm u}^2}{\omega_g^2} - 1 \right)
\ee
($m_{\rm b} \propto f,$ therefore $m_{\rm b}+m_{\rm b} \propto f^2$) which agrees with numerical calculations in Figs.~\ref{Fig-m(f)_gu}, \ref{Fig_3}.

%\noindent 
\underline{\textit{Large filling factor, $\delta f = 1-f \ll 1$.}}
In this case, the resonances (see Eq.~\eqref{Eq_w_g} in the Appendix) split into dark and bright modes. Taking, for example $N=1$, we obtain to linear order in $\delta f = 1-f$:
\be
\omega_{\rm b} = \omega_g, \quad \omega_{\rm d} = \omega_g \left[1+\delta f \left( 1- \frac{\omega_g^4}{\omega_{\rm u}^4}\right) \right].
\label{Eq-large-f}
\ee
A non-zero effective mass arises to linear order in the parameter $\delta f$:
\be
m_{\rm b} = - m_{\rm d} = \frac{\delta f}{ 2 \omega_g} \left( \frac{2 \pi}{L}\right)^2 \left(1 - \frac{\omega_{\rm u}^4}{\omega_g^4} \right) <0.
\ee
\noindent \underline{\textit{Intermediate filling factor}, $f \sim 1 .$}
From the condition $q_{\rm u} L_{\rm u} = q_{\rm g} L_{\rm g}=\pi$, using relations \eqref{q_ungated}, \eqref{q_gated} together with \eqref{Eq_wg_wu}, we find the frequency at which band degeneracy occurs
\be
\omega_0 = \frac{\omega_g}{4} \left( 1+\sqrt{1+\frac{8 \omega_{\rm u}^2}{ \omega_g^2}}\right).
\label{Eq_w0_2}
\ee
(this equation can be written in a different way, see Eq.~ \eqref{eq:w0+uslovie} in the Appendix) at a filling factor equal to
\begin{equation}
f = f^* = \frac{\omega_g^2}{4 \omega_{\rm u}^2} \left(\sqrt{1+\frac{8 \omega_{\rm u}^2}{\omega_g^2}} -1\right).
\label{Eq_fs}
\end{equation}

In the vicinity of $f\approx f^*$, we obtain the following expressions for the effective masses:
\be
m_{b} = -m_{\rm d} = - \frac{\hbar (f-f^*)}{ \omega_g} \left(\frac{2 \pi}{L}\right)^2 \frac{(2-f^*)(1-2 f^*)}{4 (1-f^*)^2 f^* }.
\ee
We see that effective mass of both bright and dark modes changes sign at the point $f =f^*$ in accordance with Figs.~\ref{Fig-m(f)_gu}, \ref{Fig_3}.

Let us now make some estimates. For $f \ll 1$ we obtain: $m_{\rm d} = -m_{\rm b} \approx 6.2 \times 10^{-4} f m_0.$ For $f \approx f^* \approx 0.28,$ we have: $m_{\rm d} = -m_{\rm b} \approx -4.8 \times 10^{-4} (f-f^*) m_0.$ Finally, for $f \approx 1,$ we get: $m_{\rm d} = -m_{\rm b} \approx -1.8 \times 10^{-4} (1-f) m_0$
where $m_0$ is the effective electron mass in GaAs.

\section{Localization of Plasmon Modes}
\label{sec-localization}
In the previous sections, we discussed how to tune the plasmon mass (or velocity, in the case of a linear spectrum) using gate voltage and filling factor $f$. In this section, we will briefly discuss how to control another important property of the plasmons: the localization of plasmonic oscillations within the crystal cell. For simplicity, we restrict our analysis to bright modes only (the analysis of dark modes is fully analogous).
As we will see, parameter $\eta$ introduced in Eq.~\eqref{eta}, along with factor $f$ and radiation frequency, plays a key role in analyzing the distribution of oscillating currents of plasma oscillations within a crystal cell.
\subsection{Gated-gated structure}
We begin by analyzing of a gated-gated system, in which $\eta$ has a very simple form (see Eq.~\eqref{Eq_eta_gg}). First, we note that for $\eta \approx 1,$ scattering at the boundaries between the different regions is very weak, so the oscillating current amplitude is approximately the same in both regions. We will therefore focus on the more interesting cases of very large and very small $\eta,$ both corresponding to $Z \gg 1,$ i.e., the strong coupling regime. 

The key ideas of plasmon localization can be illustrated with Figs.~\ref{Fig-w(f)_strong} and \ref{Fig-j(x)}.
Fig.~\ref{Fig-w(f)_strong} shows the frequency dependence of the lowest bright modes on the filling factor for $\eta=1/10.$ Red dashed lines correspond to small ($f=0.1$), intermediate ($f=0.5$), and large $f=0.9$) filling factors. The distribution of currents within the crystal cell for the four modes, indicated by the colored points in Fig.~\ref{Fig-w(f)_strong}, is displayed in the various panels of Fig.~\ref{Fig-j(x)}. This distribution was calculated using standard transfer-matrix method (see the analytical expressions for these matrices in Ref.~\cite{Gorbenko2024}).
\subsubsection{ Control of localization by frequency}
First, we consider a structure with a filling factor $f=1/2$ ($L_1=L_2=L/2$) i.e., blue and orange points in Figs.~\ref{Fig-w(f)_strong} and \ref{Fig-j(x)}. 
As shown on the upper panels of Fig.~\ref{Fig-j(x)}, the distribution of current within the crystal cell differs substantially between the blue and orange points. For the fundamental mode (orange point), the inhomogeneous part of the current is located in region ``2'', while in region ``1'', an oscillating current is approximately spatially uniform. In contrast, the mode corresponding to the blue point is almost entirely concentrated in region ``1''. Therefore, the plasmon localization can be changed by increasing the frequency.
\subsubsection{ Control of localization by gate}
 Remarkably, plasmon localization can also be controlled using a gate, which is the simplest approach experimentally.
To demonstrate this, we again consider the case $f=1/2.$
 In this case, as follows from Eq.~\eqref{eta}, the transformation
 \be
 \eta \to 1/\eta
 \ee
 is equivalent to interchanging $s_1 \leftrightarrows s_2,$ i.e. simple exchange of regions ``1'' and ``2''. Hence, changing $\eta$ 
 from a small value $\eta=1/10,$ corresponding to Figs.~\ref{Fig-w(f)_strong} and \ref{Fig-j(x)}, to a large value $\eta_1= 1/\eta=10$ at $f=1/2$ (orange and blue points in Figs.~\ref{Fig-w(f)_strong} and \ref{Fig-j(x)}) leads to a simple ``jump'' of the plasmon between the two regions of the cell. For example, for the frequency corresponding to the blue point, the region of large current shifts from region ``1'' to region ``2''. Thus, by tuning the parameter $\eta$ via gate voltages, the localization of plasmons can be readily altered.
\subsubsection{Control of localization by filling factor}
The localization of plasmons also depends significantly on the factor $f.$ To illustrate this, consider three points on the lower curve of Fig.~\ref{Fig-w(f)_strong}. As seen from Fig.~\ref{Fig-j(x)}, for small $f$ (green point), similar to the case of intermediate $f$ (orange point), the plasmon (more precisely, the inhomogeneous part of the alternating current) is localized in region ``2''. However, by increasing the filling factor to values close to unity, the plasmon can be ``transferred'' to region ``1'', as illustrated on the right lower panel of Fig.~\ref{Fig-j(x)} corresponding to the purple point with $f=0.9.$

\subsection{Gated-ungated structure}
The switching of localization regions is also possible in a gated-ungated structure, which is significantly easier to implement experimentally. In this case, assuming that the gate is over region ``1'' and region ``2'' is open, the expression for $\eta$ is given by formula \eqref{Eq_eta_gu}. Similar to gated-gated case, the switching of localization region occurs with increasing frequency at fixed and $f,$ or with changing $\eta$ from small to large values via the gate.

\begin{figure}[h!]
\centering
\includegraphics[width=6.6 cm]{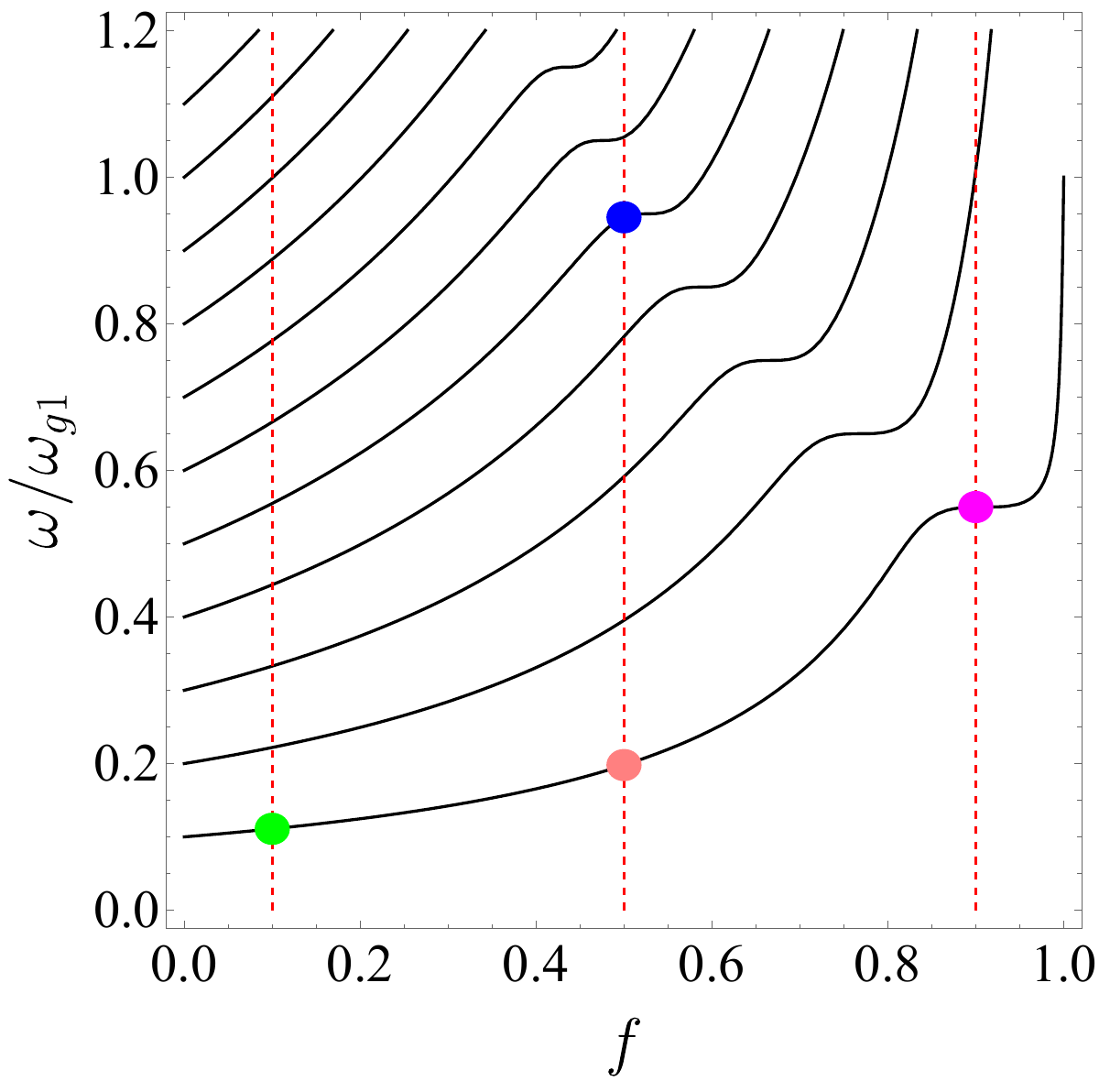}
\caption{Dependencies of the eigenfrequencies of the plasmonic crystal at $q=0$ in units of $\omega_{\rm g1}= 2\pi s_1/L$ on the filling factor for a gated-gated structure with a small parameter $\eta$: $s_1 = 10 s_2,$ $\eta =1/10$. Dashed lines indicate $f=0.1, 0.5, 0.9$.}
\label{Fig-w(f)_strong}
\end{figure}

\begin{figure}[h!]
\centering
\includegraphics[width=8.6 cm]{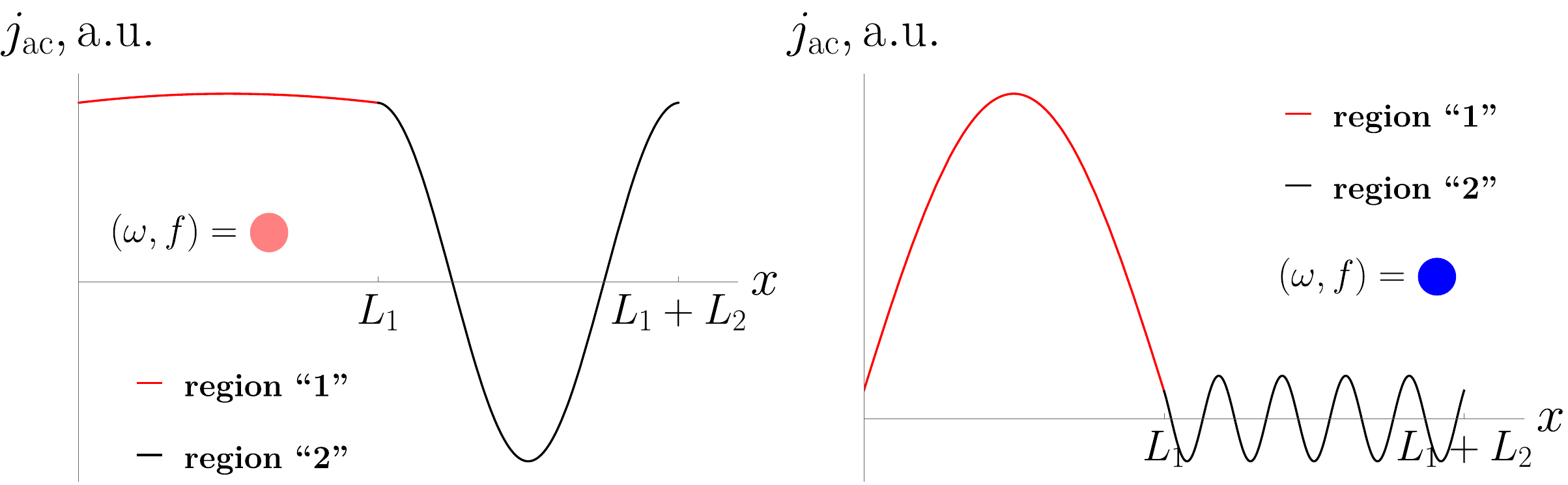}

\includegraphics[width=8.6 cm]{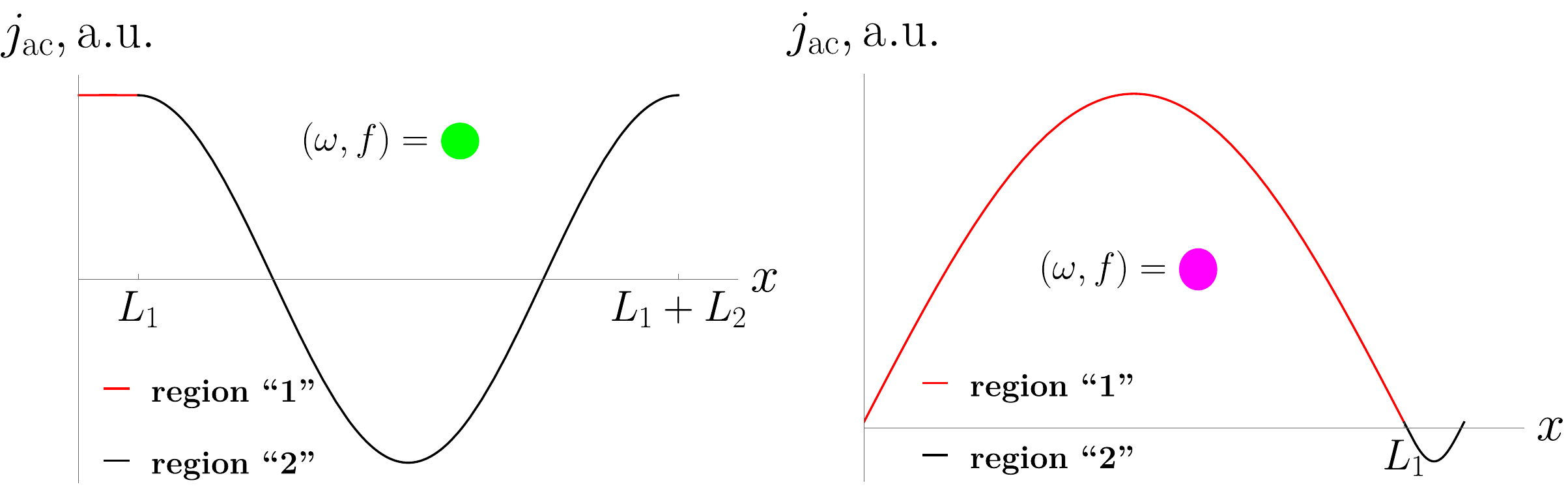}% Here is how to import EPS art
\caption{Distributions of plasmonic ac currents at various resonant frequencies of the plasmonic crystal in a gated-gated structure. Colored thick dots correspond to the ones in Fig.~\ref{Fig-w(f)_strong}}
\label{Fig-j(x)}
\end{figure}

In experiments \cite{Sai2023}, performed on a number of gated-ungated structures with $f \sim 1$ (not too small and not too close to unity) a potential was applied to the grating gate to deplete electrons in the gated region, i.e., $s_1$ was very small, so that for all experimental frequencies, the condition $s_1 \ll \omega h_1$ was satisfied, and consequently, the inequality $\eta \gg 1$ held. Frequency in this experiment was higher than fundamental frequency. Hence, this case corresponds to the one described by blue point in Figs.~\ref{Fig-w(f)_strong} and \ref{Fig-j(x)}. Given that $\eta$ was large, we conclude that the plasmon was localized in the ungated region ``2''.

Conversely, in the experiment \cite{Khisameeva2025}, where no voltage was applied to the gate (i.e., the condition $N_{\rm u}= N_{\rm g}$ was satisfied), the parameter $\eta$ was less than unity ($ \eta \approx 0.3-0.4$),
 the fundamental plasmon at $f\approx 1/2$ was concentrated in the gated region (see Fig.~ S4 in Ref.~\cite{Khisameeva2025}).

\section{Conclusion}

Massive bright and dark plasmonic excitations in plasma crystals of various geometries were studied. It was demonstrated that the effective masses of both plasmons can be controlled by applying gate voltages. An acoustic plasmon mode existing at low frequencies was also investigated, as well as additional modes with a linear dispersion that emerge at specific gate voltages. It was shown that the spatial localization of plasmons can also be modified using gate voltages. Specifically, in a gated-ungated structure, at zero gate voltage, the plasmon is concentrated in the gated region. When a potential is applied that depletes electrons from the gated region, the plasmon shifts to the ungated region. The results of this study provide a foundation for the efficient band-gap engineering of plasmonic crystals, enabling purely electrical control of the key parameters of plasma excitations.

\section*{Acknowledgments}
The work of I.G. and V.K. was supported by the 
Russian Science Foundation under grant 24-62-00010. The work of P.G. and V.M. was supported as a part of the ISSP RAS State assignment. The work of I.G. was also partially supported by the Theoretical Physics and Mathematics Advancement Foundation ``BASIS''.

\appendix

\section{Degeneracy Frequencies}\label{sec_modes_degeneracy}
In this Appendix, we analyze several cases in which the degeneracy $\omega_{\rm b}=\omega_{\rm d}=\omega_0$ occurs and demonstrate that in the degenerate case, the crystal's response to an external field is significantly simplified compared to the general case.

\subsection{Various Cases of Degeneracy}
Degeneracy occurs due to several different reasons:

\textbf{1)} The first case follows explicitly from the fact that $B(\omega)$ and $D(\omega)$ differ by the replacement $\eta \to 1/\eta$. For $\eta = 1$, the expressions are identical: $B(\omega) = D(\omega) = \sin{(q_1 L_1+q_2 L_2)/2}$.
In the gated-gated case, this condition (see Eq.~\eqref{Eq_eta_gg}) means $s_1=s_2,$ i.e., homogeneity of the electron fluid.
At the same time, in a gated-ungated structure, the case $\eta=1$ is realized even in an inhomogeneous electron fluid. From Eq.~\eqref{Eq_eta_gu} it follows that $\eta=1$ at the frequency
\be
\omega_0 = \frac{s_1}{2 h} \frac{\varepsilon_{\rm eff}}{\varepsilon}.
\ee
For a given geometry $h, L_1, L_2$, this frequency can be controlled by the voltage on the grating gate, by changing $s_1$.
Since $h \ll L,$ this frequency is quite high for the case when no gate voltage is applied and $N_{\rm g} =N_{\rm u}.$ However, by applying a ``depleting'' voltage to the gate, one can lower $s_1$ and, consequently, the frequency $\omega_0$. For example, in Figure \ref{Fig-w(f)_gu}, corresponding to the case of strong depletion, $N_{\rm g} \ll N_{\rm u}$, this frequency is shown by a dashed line.

\textbf{2)} The limiting cases of zero ($f=0$) and full ($f=1$) filling also lead to the degeneracy of all dark and bright frequencies, regardless of other parameters. This is clearly seen from Figures \ref{Fig-m(f)_gg} and \ref{Fig-m(f)_gu}, where the cases $f=0$ and $f=1$ correspond to zero plasmon masses. Indeed, for $L_1 \to 0$: $B(\omega) = D(\omega) = \sin{q_2 L_2/2}$, and similarly, for $L_2 \to 0$, $B(\omega) = D(\omega) = \sin{q_1 L_1/2}$. In both cases, all eigenfrequencies are degenerate: $\omega_{\rm b} = \omega_{\rm d} = \omega_0$.

It is convenient to rewrite the expression for $\omega_0$ by introducing the frequencies $\omega_{\rm u}$ and $\omega_{\rm g}$ (see Eq.~\eqref{Eq_wg_wu}).
In the gated-ungated system, the case $L_1 \to 0$ corresponds to a zero filling factor, $f=0$, and the frequencies at which degeneracy occurs are given by the expression
$
\omega_0 = N \omega_{\rm u},
\label{Eq_w_u}
$
where $N$ is an integer.
The case of complete screening by the gate corresponds to $f=1$ ($L_2 \to 0$), and the degeneracy frequencies are
\begin{equation}
\omega_0 = N \omega_{\rm g}.
\label{Eq_w_g}
\end{equation}
In the gated-gated case, the limit $L_1 \to 0$ corresponds to frequencies
$
\omega_0 = N \omega_{\rm g 2} $
and, similarly, for $L_2 \to 0$, to frequencies
$
\omega_0 = N \omega_{\rm g 1}.
$

\textbf{3)} The third and most interesting case of degeneracy, which is also the most experimentally accessible, occurs when the phases of plasma waves in the two regions satisfy the following condition:
\begin{equation}
q_1 L_1 = \pi N, \quad q_2 L_2 = \pi M,
\label{Eq_case2}
\end{equation}
with the integers $N$ and $M$ having the same parity (see a more detailed discussion in section III.D of Ref.~\cite{Gorbenko2024}). This condition nullifies both terms in $B(\omega)$ as well as in $D(\omega)$, regardless of the parameter $\eta$ (see Eq.~\eqref{Eq_BD}), and is universal for both gated-gated and gated-ungated systems.

In a gated-gated structure, the degeneracy condition yields
\begin{equation}
\omega_0=\frac{\pi N s_{\rm 1}}{L_{\rm 1}}=\frac{\pi M s_{\rm 2}}{L_{\rm 2}}.
\end{equation}
Obviously, for given $N,M$ and $L_{1,2},$ the second of these equalities can be satisfied only for specific values of gate voltages, corresponding to the condition
\begin{equation}
\frac{s_2}{s_1}= \frac{N L_2}{M L_1}.
\end{equation}
(see also equations (34) and (35) in work \cite{Gorbenko2024}).

In a gated-ungated structure, the degeneracy frequency can be written in two equivalent forms
\begin{equation}
\omega_0=\frac{\pi N s_{\rm g} }{L_{\rm g}}=\sqrt{\frac{2\pi^2 e^2 N_{\rm u} M }{m \varepsilon_{\rm eff} L_{\rm u}} }.
\label{eq:w0+uslovie}
\end{equation}
The second of these equalities gives the condition
\begin{equation}
\frac{2\pi N^2}{M} \frac{N_{\rm g}}{N_{\rm u}}=\frac{\varepsilon}{\varepsilon_{\rm eff}} \frac{L_{\rm g}^2}{ L_{\rm u} h},
\end{equation}
which can be satisfied by changing the concentration $N_{\rm g}$ by using the gate.

For the degeneracy of the fundamental bright and fundamental dark mode, i.e. for $M=N=1,$ the following conditions should be satisfied
\be
q_1 L_1=q_2 L_2=\pi. 
\ee

In the vicinity of the corresponding frequency, $\omega_0=\omega_{\rm b} = \omega_{\rm d} ,$ the spectrum is linear and characterized by a certain velocity $s_0$, which is different for gated-gated and gated-ungated structures. Analytical expressions for these velocities are given in the main text of the article, see Eq.~\eqref{eq-s0}.

\subsection{Response to an External Field at Frequency $\omega_0$}
In Ref.~\cite{Gorbenko2024}, a general expression for the transmission coefficient through a gated-gated structure for arbitrary values of parameters $\eta$ and $f$ was derived (see Eqs.~(1) and (21) in Ref.~ \cite{Gorbenko2024}). For the degenerate case, i.e., under the condition $q_1 L_1 = q_2 L_2$, the analytical expression for the transmission coefficient through the plasmonic crystal is significantly simplified. For simplicity, we will consider the gated-gated case only. Then, the condition $q_1 L_1 = q_2 L_2$ means $L_2/s_2 = L_1/s_1$ and introducing the notations $f=L_1/L$, $\omega_0 = \pi s_1/L_1 = \pi s_2/L_2$, we obtain for the transmission coefficient through the plasmonic crystal $T=1-\delta T$:
\begin{equation}
\begin{aligned}
&\delta T = \frac{4 L^2 \omega_0^2 }{\pi c \sqrt{\varepsilon}}\frac{C_0 \gamma}{ (\omega^2+\gamma^2)}
\\
&\times \left(1-3f+3 f^2+\frac{2 (1-f)^2 \omega_0 }{ \pi \gamma (\Omega^2+\Gamma^2) (\cos{\frac{\pi \Omega }{\omega_0}}+\cosh{\frac{\pi \Gamma}{\omega_0}})} \right.
\\
&\left. \times \left[(\omega \Omega - \gamma \Gamma) \sinh{\frac{\pi \Gamma}{\omega_0}} -(\omega \Gamma + \gamma \Omega) \sin{\frac{\pi \Omega}{\omega_0}} \right] \right)
\end{aligned}
\end{equation}
where $c$ is the speed of light, $C_0 = \varepsilon/4 \pi h$ is the channel capacitance per unit area, $\gamma=1/\tau$ is the inverse momentum relaxation time, $\Omega = {\rm Re}[\sqrt{\omega(\omega+i\gamma)}]$, $\Gamma = {\rm Im}[\sqrt{\omega(\omega+i\gamma)}]$. Here, for simplicity, we assume that the dielectric permittivity is the same everywhere and equal to $\varepsilon.$

\bibliography{main.bib}

\end{document}